\documentclass[aps, preprint,prl, superscriptaddress]{revtex4-1}
\usepackage{titlesec}%
\usepackage{subfigure}%
\usepackage{amsmath}%
\usepackage{amsfonts}%
\usepackage{amssymb}%
\usepackage{feynmp}%
\usepackage{graphicx}%
\usepackage{setspace}%
\usepackage{comment}%
\usepackage{dsfont}%
\usepackage{color}%
\usepackage{bbold}
\usepackage{tikz}
\usetikzlibrary{arrows,calc}
\usepackage{hyperref}
\hypersetup{colorlinks,allcolors=black}

\usepackage{xspace}
\usepackage{siunitx}
\usepackage{lineno}

\titleformat{\paragraph}
{\normalfont\normalsize\bfseries}{\theparagraph}{1em}{}
\titlespacing*{\paragraph}
{0pt}{3.25ex plus 1ex minus .2ex}{1.5ex plus .2ex}

\begin{document}

\title{Two-photon Interference of Biphotons Emitted by Overlapping Resonances in Metasurfaces}

\author{Jiho Noh}
\affiliation{Sandia National Laboratories, Albuquerque, New Mexico 87185, USA.}
\affiliation{Center for Integrated Nanotechnologies, Sandia National Laboratories, Albuquerque, New Mexico 87185, USA.}

\author{Tomás Santiago-Cruz}
\affiliation{Sandia National Laboratories, Albuquerque, New Mexico 87185, USA.}
\affiliation{Center for Integrated Nanotechnologies, Sandia National Laboratories, Albuquerque, New Mexico 87185, USA.}

\author{Chloe F. Doiron}
\affiliation{Sandia National Laboratories, Albuquerque, New Mexico 87185, USA.}
\affiliation{Center for Integrated Nanotechnologies, Sandia National Laboratories, Albuquerque, New Mexico 87185, USA.}

\author{Hyunseung Jung}
\affiliation{Sandia National Laboratories, Albuquerque, New Mexico 87185, USA.}
\affiliation{Center for Integrated Nanotechnologies, Sandia National Laboratories, Albuquerque, New Mexico 87185, USA.}

\author{Jaeyeon Yu}
\affiliation{Sandia National Laboratories, Albuquerque, New Mexico 87185, USA.}
\affiliation{Center for Integrated Nanotechnologies, Sandia National Laboratories, Albuquerque, New Mexico 87185, USA.}

\author{Sadhvikas J. Addamane}
\affiliation{Sandia National Laboratories, Albuquerque, New Mexico 87185, USA.}
\affiliation{Center for Integrated Nanotechnologies, Sandia National Laboratories, Albuquerque, New Mexico 87185, USA.}

\author{Maria V. Chekhova}
\affiliation{Max Planck Institute for the Science of Light, 91058 Erlangen, Germany.}
\affiliation{Friedrich-Alexander-Universität Erlangen-Nürnberg, 91058 Erlangen, Germany.}

\author{Igal Brener}
\affiliation{Sandia National Laboratories, Albuquerque, New Mexico 87185, USA.}
\affiliation{Center for Integrated Nanotechnologies, Sandia National Laboratories, Albuquerque, New Mexico 87185, USA.}
\email{ibrener@sandia.gov}

\date{\today}

\maketitle

\textbf{Abstract}

Two-photon interference, a quantum phenomenon arising from the principle of indistinguishability, is a powerful tool for quantum state engineering and plays a fundamental role in various quantum technologies.
These technologies demand robust and efficient sources of quantum light, as well as scalable, integrable and multifunctional platforms.
In this regard, quantum optical metasurfaces (QOMs) are emerging as promising platforms for quantum light generation, namely biphotons via spontaneous parametric down-conversion (SPDC), and its engineering.
Due to the relaxation of phase matching, SPDC in QOMs allows different channels of biphoton generation, such as those supported by overlapping resonances, to occur simultaneously.
In previously reported QOMs, however, SPDC was too weak to observe such effects. Here we develop QOMs based on [110]-oriented GaAs that provide more than an order of magnitude enhancement in SPDC rate, after accounting for the spectral bandwidth, compared to any other QOMs studied to date. 
This boosted efficiency allows the QOMs support the simultaneous generation of SPDC from several spectrally overlapping optical modes.
Using polarization components in the interferometer analyzer, we intentionally erase the distinguishability between the biphotons from a high-$Q$ quasi-bound-state-in-the-continuum resonance and a low-$Q$ Mie resonance, which results in the first-time observation of two-photon interference in the spectral domain in these types of devices.
This quantum interference can considerably enrich the generation of entangled photons in metasurfaces. Their advanced multifunctionality, improved nonlinear response, ease of fabrication and compact footprint of [110]-GaAs QOMs position them as promising platforms to fulfill the requirements for photonic quantum technologies.

\newpage

Quantum interference is an interesting phenomenon that has facilitated many advances in
quantum technologies, such as in quantum information processing~\cite{Stobinska_SciAdv_2019,Qiang_NatPhoton_2018}, quantum computation ~\cite{Knill_Nature_2001,Spring_Science_2013,Tillmann_NatPhoton_2013}, quantum networks~\cite{Bao_NatPhoton_2023}, quantum imaging, sensing, and metrology~\cite{Lemos_Nature_2014,Paterova_SciAdv_2020,Kviatkovsky_SciAdv_2020,Clark_ApplPhysLett_2021,Defienne_NatPhoton_2024}. 
Much of this progress is enabled by the use of entangled two-photon (biphoton) light, which can be generated by nonlinear processes such as spontaneous parametric down-conversion (SPDC) or spontaneous four-wave mixing in second- and third-order nonlinear materials, respectively. A quantum-mechanical effect that sparked considerable discussion in the 1990s (see, for instance, Ref.~\cite{Mandel_RevModPhys_1999}) is two-photon interference, a.k.a fourth-order interference, between biphotons generated in two \textit{spatially separated} nonlinear materials.
Specifically, when two nonlinear materials are pumped coherently by the same laser beam, the biphotons emitted from the two sources may exhibit two-photon interference if they are indistinguishable --even in principle--~\cite{Feynman1965}.
This phenomenon was utilized to generate polarization-entangled photons~\cite{Kwiat_PhysRevA_1999,Burlakov_PRA_2001, Brida_PRL_2006} and is used in modern implementations of quantum nonlinear interferometers~\cite{Chekhova_AdvOptPhoton_2016}.

In this work, we introduce a unique platform to observe two-photon interference in the spectral domain between biphotons emitted from two \textit{spatially overlapping yet distinct} sources in the same nonlinear medium.
Our approach exploits SPDC from so-called `quantum optical' metasurfaces (QOMs). 
Metasurfaces made from nonlinear materials have already revolutionized classical nonlinear optics~\cite{Chen_RepProgPhys_2016, Li_NatRevMats_2017,Krasnok_MaterToday_2018, Sain_AdvPhoton_2019,Vabishchevich_PhotonRes_2023}, and they recently have emerged as a promising platform for addressing the challenges of engineering the quantum state of biphotons.
These nanostructures, carefully designed with features smaller than the wavelength of light, dramatically alter how light interacts with the material. By manipulating the phase, amplitude, and polarization of light at the nanoscale, metasurfaces enable tailored light-matter interactions far surpassing those achievable with conventional methods~\cite{Chen_RepProgPhys_2016, Li_NatRevMats_2017, Sain_AdvPhoton_2019,Vabishchevich_PhotonRes_2023}. 
The natural extension of metasurfaces into the quantum regime, leading to QOMs, promises unprecedented control over biphoton generation, potentially overcoming the limitations inherent in biphotons generated in bulk materials.
A groundbreaking demonstration of SPDC from LiNbO$_{3}$ QOMs in 2021~\cite{Santiago-Cruz_NanoLett_2021} showcased the feasibility of this approach, catalyzing a series of research exploring various material systems and design strategies. 
Since then, researchers successfully developed SPDC sources exhibiting diverse features, including frequency multiplexing~\cite{Santiago-Cruz_Science_2022}, spatial entanglement~\cite{Zhang_SciAdv_2022}, bidirectional emission~\cite{Son_Nanoscale_2023,Weissflog_Nanophotonics_2024}, polarization entanglement along with Bell-state generation~\cite{Jia_arXiv_2024,Ma_arxiv_2024b}, and high-dimensional entanglement through multiphoton-state generation~\cite{Li_Science_2020}.

To observe two-photon interference, we designed metasurfaces supporting a high-quality ($Q$) factor resonance, namely a quasi-bound state in the continuum (qBIC), along with a low-$Q$ in-plane Mie-like resonance, similar to our previous works~\cite{Santiago-Cruz_Science_2022,Noh_NanoLett_2024}.
Unlike the aforementioned works where the in-plane Mie-like mode did not make a significant contribution to SPDC primarily due to the unfavorable crystal orientation of the metasurface’s material -- [001]-oriented GaAs -- here, we dramatically increase the biphoton generation rate.
To this end, we fabricate the metasurfaces from a [110]-oriented GaAs wafer, whose crystal orientation simultaneously boosts the SPDC emission driven by Mie-type~\cite{Ma_arxiv_2024b} and qBIC resonances.
The transition from the conventional [001] orientation to the [110] orientation of GaAs improves the overlap integral between the electric field of interacting modes and the second-order nonlinear tensor of GaAs, thereby amplifying the efficiency of SPDC and its classical reverse process, second harmonic generation (SHG). 
As we will show below, this strategy results in biphoton-generation rates that surpass that of any previously reported QOM by more than an order of magnitude. At the same time, the enhanced efficiency leads to the first observation of quantum interference phenomena in QOMs.
This strategy pushes forward the boundaries of what is possible with QOMs without compromising their multifunctionality.

\begin{figure}[t!]
\centering
\includegraphics[width=14cm]{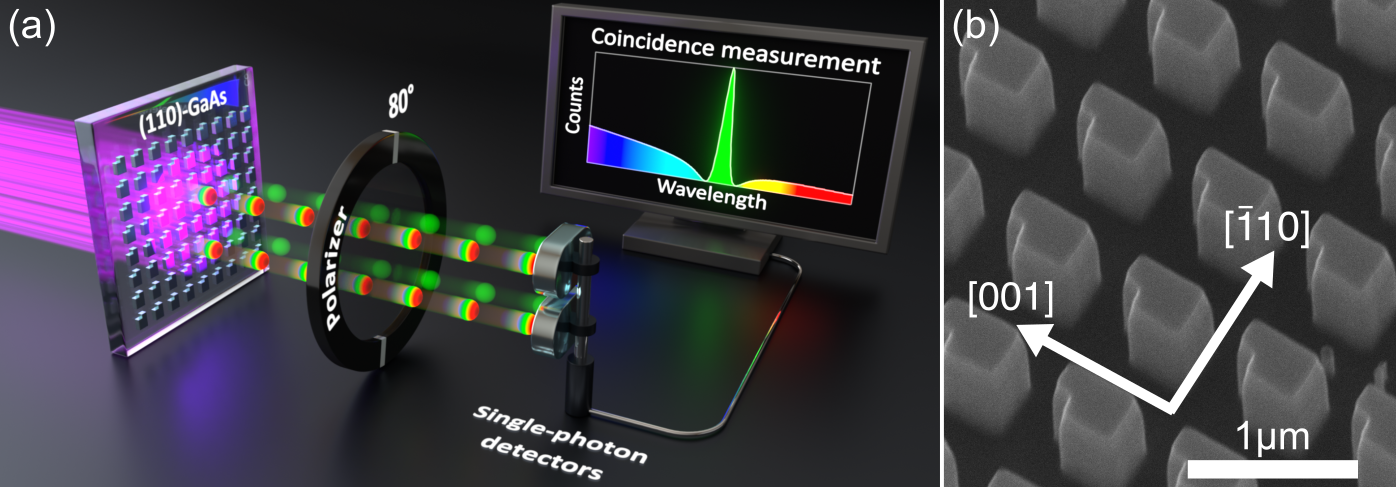}
\caption{(a) Schematic diagram of two-photon interference of biphotons emitted by spatially overlapping yet distinct resonances in [110]-GaAs metasurfaces. (b) Scanning electron microscopy (SEM) image of the metasurface.}
\label{fig:Intro}
\end{figure}

Materials with zinc-blende crystalline structure such as GaAs, AlGaAs, GaP, InGaP, etc, are particularly attractive for nonlinear wave mixing because they feature some of the highest second-order susceptibilities~\cite{Shoji_JOSAB_1997}, but their standard crystalline orientation -- [001] -- commonly used in the fabrication of resonant platforms, is not optimal for this purpose. 
Indeed, it has been predicted that metasurfaces fabricated on [110]- and [111]-oriented GaAs may exhibit a stronger second-order nonlinear response than those fabricated on [001]-oriented GaAs~\cite{Liu_NatCommun_2018}.
This enhancement is attributed mainly to a better overlap integral between the $\chi^{(2)}$ tensor and the electric field profiles of the interacting modes. 
Meanwhile, some experimental studies have demonstrated the potential advantages of specific GaAs crystallographic orientations for nonlinear optical processes.
For instance, [110] and [111] crystallographic orientations of GaAs have been shown to facilitate normal emission of SHG and its steering in the forward or backward directions in single nanoantennas~\cite{Sautter_NanoLett_2019,Xu_ACSNano_2020}.
Motivated by these findings, R. Camacho-Morales et al. used [110]-GaAs metasurfaces to enhance the nonlinear mixing of two co-propagating beams~\cite{Camacho-Morales_AdvPhotonics_2021}, while M. Yang et al. recently demonstrated strong SHG emission in the zeroth diffraction order in monolithic [110]-GaP metasurfaces~\cite{Yang_Nanophotonics_2024}.
However, to the best of our knowledge, no single study has experimentally benchmarked the performance of [110] or [111] metasurfaces against their [001] counterpart in either SHG or SPDC.
Moreover, the aforementioned works exploited only low-$Q$ in-plane dipole resonances, while in this work we use out-of-plane modes with different symmetries and much higher $Q$.

Our metasurface design exploits the concept of qBICs, beginning with an array of square meta-atoms that exhibit $C_{4v}$ symmetry,  supporting symmetry-protected BICs with infinite $Q$~\cite{Hsu_NatMat_2016, Koshelev_PRL_2018}.
By carefully breaking this symmetry, we transform these BICs into qBICs that maintain high $Q$-factors while becoming experimentally accessible due to weak coupling to far-fields~\cite{Campione_ACSPhoton_2016,Vabishchevich_ACSPhotonics_2018}.
These metasurfaces support both out-of-plane electric dipole (ED) and magnetic dipole (MD) qBICs~\cite{Santiago-Cruz_Science_2022}.
Along with qBICs, the metasurfaces also support a set of in-plane Mie-type dipole modes that stem from the initial square shape of the meta-atoms~\cite{Campione_ACSPhoton_2016, Noh_NanoLett_2024}.
At resonance wavelengths, both qBICs, and Mie-type modes enhance the zero-point vacuum fluctuations that seed SPDC.
Due to their higher $Q$ factors, qBICs enhance the biphoton generation rate spectral density more significantly than low-$Q$ in-plane Mie modes~\cite{Santiago-Cruz_NanoLett_2021}.

\begin{figure}[t!]
\centering
\includegraphics[width=12cm]{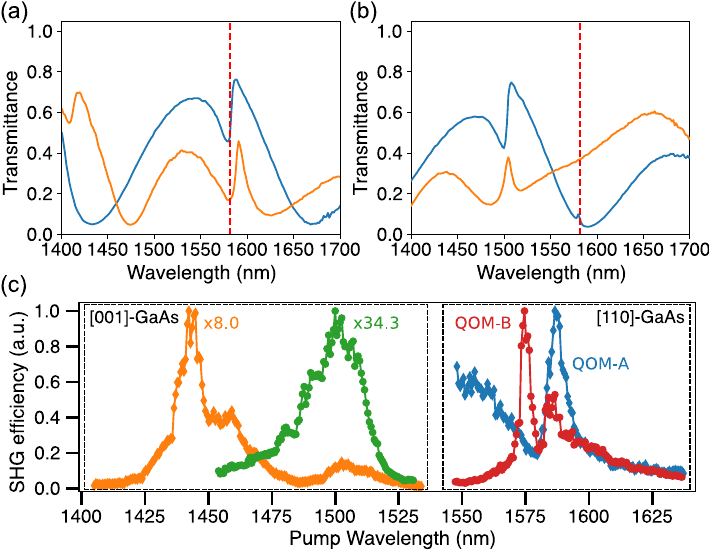}
\caption{(a) Measured white-light transmission spectra of QOM-A and (b) QOM-B for incident polarizations along the direction 45$^{\circ}$ tilted from [$\bar{1}$10] toward [001] (blue) and its orthogonal direction (orange). Red dashed line indicates double the wavelength of the pump beam. (c) SHG spectroscopy in [001]- (left dashed box) and [110]-oriented (right dashed box) GaAs metasurfaces. The diamonds and circles show the SHG efficiency from ED-qBIC and MD-qBIC resonances, respectively. At the ED-qBIC resonance, the SHG efficiency is 8-fold enhanced in [110]-GaAs metasurfaces due to an improved mode overlap. The effect of the MD-qBIC resonance in the [001]-GaAs metasurface (green circles) was not observable because the fs laser was spectrally too broad to couple to the resonance ($Q\sim$ 1590).}
\label{fig:Spectra}
\end{figure}

We fabricated two types of [110]-GaAs QOMs, QOM-A, and QOM-B, featuring ED-qBIC at 1588~nm and MD-qBIC at 1579~nm, respectively, to generate frequency-degenerate biphotons from our 790.8~nm continuous-wave (cw) laser, as presented in Figures~\ref{fig:Spectra}(a) and (b).
In addition, with the same laser, the in-plane Mie modes spanning a larger wavelength range will emit frequency-nondegenerate biphotons.
The measured $Q$-factors of MD- and ED-qBICs were around $10^{2}$, as in the previous work for SPDC with [001]-GaAs metasurface.~\cite{Santiago-Cruz_Science_2022}
To fabricate the QOMs, we utilized a 500~nm-thick [110]-oriented GaAs film grown by molecular beam epitaxy, off-orientated by 6$^{\circ}$ towards [111]A to improve surface quality.
The details of the fabrication process, illustrated in Figure~\ref{fig:FabProcess}, is in supporting information (SI); the metasurface dimensions are shown in Table~\ref{tab:dimensions}.

First, we experimentally verify that SHG is enhanced in our [110]-GaAs metasurfaces. We pump the metasurfaces from the air side with a pulsed laser (350 fs, 1 MHz) in a custom-built SHG spectroscopy setup as shown in Figure~\ref{fig:setup}(a).
The linear polarization of the pump laser was adjusted for each metasurface to optimize the nonlinear interaction. Figure~\ref{fig:Spectra}(c) shows the measured SHG efficiency ($P_{SH}/P^{2}_{pump}$) as a function of wavelength for the ED-qBIC (diamonds) and MD-qBIC (circles) resonances in QOM-A and QOM-B, respectively (right dashed box). 
The results show that SHG peaks at the resonance wavelengths of ED-qBIC and MD-qBIC modes. Although the MD-qBIC features a higher $Q$-factor than that of the ED-qBIC, their SHG responses are identical. This minor discrepancy may arise from various factors, including a weaker mode overlap between the electric field of the MD-qBIC mode and the $\chi^{(2)}$ tensor, as well as a non-optimal coupling of our Gaussian pump beam to the MD-qBIC mode.
In QOM-A, we can also observe SHG contributions from a low-$Q$ in-plane Mie mode, represented by the decreasing slope between 1550 nm and 1575 nm and a Fano-type dip at 1580~nm.

For benchmarking, we further tested, under identical experimental conditions, the SHG of an ED-qBIC resonance at 1446.9 nm ($Q$ $\sim$ 330) from an [001]-GaAs metasurface (orange diamonds in Figure~\ref{fig:Spectra}(c)), which was used in our previous work for SPDC~\cite{Santiago-Cruz_Science_2022}.
Within the effective collection numerical aperture (NA) of our setup (NA $\sim 0.16$), the SHG is 8-fold enhanced in both QOM-A and QOM-B compared to that of the ED-qBIC resonance in the [001]-GaAs metasurface.
We therefore expect similar performance in SPDC.
The same [001]-GaAs metasurface exhibits an MD-qBIC at 1511.8~nm with a higher $Q$-factor ($Q\sim$ 1590), but our fs laser is spectrally too broad to efficiently excite this resonance.
Notably, the response of the low-$Q$ in-plane Mie mode in the [001]-GaAs metasurface is much weaker than in the [110]-GaAs metasurface, which is reflected in the weak and broad response indicated by green circles in Figure~\ref{fig:Intro}(c).
To confirm the enhanced SHG efficiencies in [110]-oriented GaAs metasurfaces, we performed SHG simulations using the COMSOL Multiphysics frequency-domain finite element method solver, with the detailed methods provided in the SI. 
Under optimum conditions, where both the pump beam polarization and the metasurface orientation with respect to the crystalline orientations were chosen to maximize the overlap integral between the $\chi^{(2)}$ tensor and the field profiles, the simulated SHG responses of ED-qBIC (QOM-A) and MD-qBIC (QOM-B) in [110]-GaAs were 23-fold and 11-fold higher, respectively, compared to the corresponding optimum results in [001]-GaAs, when integrated over the same NA as in the experiment. 
These results confirm our initial hypothesis on the enhanced performance of [110]-oriented GaAs metasurfaces.

Next, we focus on the main subject of this work which is SPDC and two-photon interference. 
To pump SPDC, we employ a 55~mW cw laser at $\lambda =$ 790.8~nm.
We control the linear polarization of the pump beam with a half-wave plate (HWP) and focus the beam (200~$\mu$m full width at half-maximum diameter) onto the metasurface using a 60 mm focal length lens.
A lens with a focal length of 18.4~mm collects the generated biphotons, which are then isolated from the pump beam through a cascade of long-pass filters with cut-on wavelengths at 1000~nm, 1300~nm, 1350~nm, 1400~nm, and 1450~nm.
The filtered photons are then directed to a Hanbury Brown-Twiss-like setup, consisting of a 50:50 non-polarizing beam-splitter (NPBS) cube and superconducting nanowire single-photon detectors (SNSPDs).
A time tagger records photodetection pulses from individual SNSPDs, and it tallies joint detection events by analyzing time intervals.
This allows the measurement of the rate of simultaneous photon detections, or coincidences. 
For the detection using SNSPDs, we couple the photons into single-mode fibers (SMF-28) using 18.4~mm focal length lenses, resulting in an effective collection NA of $\sim$ 0.14.
The schematic of the experimental setup is shown in Figure~\ref{fig:setup}(b).

\begin{figure}[t!]
\centering
\includegraphics[width=\textwidth]{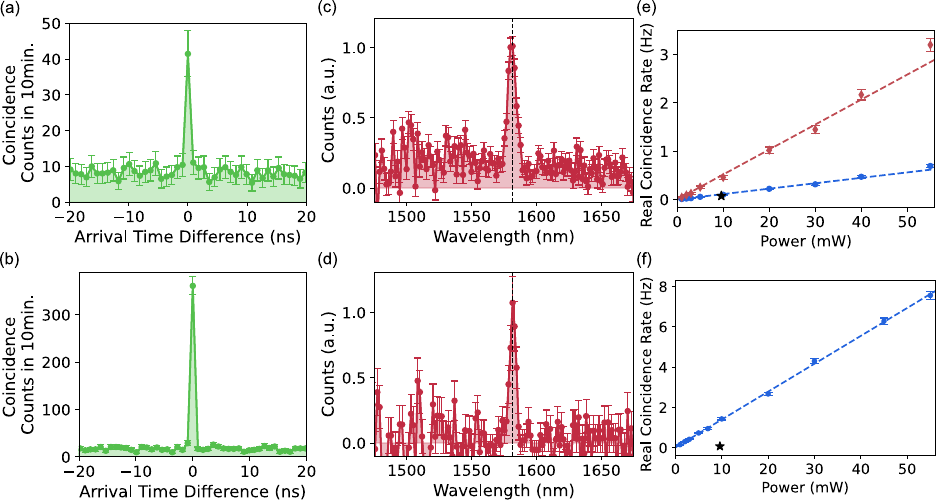}
\caption{(a-b) Coincidence histograms of SPDC from QOM-A and QOM-B, respectively. In both measurements, the pump power is 5~mW and the acquisition time is 10 min. (c-d) Measured SPDC spectra from QOM-A and QOM-B, respectively. Black dashed lines indicate double the wavelength of the pump laser, and their overlap with the peak in the SPDC spectra indicates that the SPDC processes are frequency-degenerate. The broad pedestal in panel (c) is due to non-degenerate SPDC driven by the in-plane Mie mode. (e-f) Measured power dependence of the SPDC rate from QOM-A and QOM-B, respectively. In panel (e), the red diamonds include the effect of both ED-qBIC and in-plane Mie resonances. To obtain the real coincidence rate from the ED-qBIC (blue circles), we subtracted the contribution of the in-plane Mie mode based on the data in panel (c). Error bars indicate the statistical uncertainty. Stars in (e-f) indicate previously reported SPDC rates in [001]-GaAs metasurfaces~\cite{Santiago-Cruz_Science_2022}.} 
\label{fig:SPDC}
\end{figure}

We perform coincidence measurements by pumping the metasurfaces from the substrate side and collecting the biphotons from the air side (see Figure~\ref{fig:Intro}(a)).
Figures \ref{fig:SPDC}(a) and (b) show coincidence histograms acquired from QOM-A and QOM-B, respectively, for 10~minutes and when pumping with 5~mW.
The combined timing jitter of the detectors is approximately 162~ps, and the time bin in the time tagger was set to 900~ps.
The peak-to-background ratio, i.e., the second-order correlation function $g^{(2)}(0)$, exceeds two in both cases, confirming the detection of biphotons.
Given that the metasurfaces were pumped above the bandgap of GaAs, incoherent photoluminescence (PL) emission accompanied SPDC, with the former being significantly more efficient than the latter.
The prominent background (accidental coincidence counts) in Figs. \ref{fig:SPDC}(a) and (b) is attributed solely to uncorrelated coincidence events originating from PL emission.

To verify that the coincidence events observed in Figures~\ref{fig:SPDC}(a, b) stem from the resonant behavior of the metasurfaces, we examine the emission spectrum via time-of-flight spectroscopy\cite{Valencia_PRL_2002}.
Inserting 2 km-spools of single-mode fibers (SMF-28) on each detection arm (see experimental setup in Figure~\ref{fig:setup}(b)) broadened the coincidence histograms in arrival time difference.
We then map the arrival time differences to photon wavelengths using calibration curves.
The spectral resolution, limited by the timing jitter of the detectors, was measured to be 4.3~nm.

The spectrum measured from QOM-A (Figure~\ref{fig:SPDC}(c)) shows a distinct peak with a full width at half maximum (FWHM) of 6.4~nm at $\sim$1580.9~nm atop a broadband pedestal.
The peak originates from frequency-degenerate SPDC emission driven by the ED-qBIC resonance. This behavior is expected, as the ED-qBIC resonance wavelength closely matches twice the pump wavelength, that is, energy conservation is fulfilled when the signal and idler photons have the same wavelength.~\cite{Santiago-Cruz_Science_2022}.
The pedestal, on the other hand, arises from frequency-nondegenerate SPDC from low-$Q$ in-plane Mie modes in the vicinity. Remarkably, the effect of these low-$Q$ in-plane Mie modes as the sole source for SPDC was not observed in our previous works in [001]-oriented GaAs metasurfaces,~\cite{Santiago-Cruz_Science_2022,Noh_NanoLett_2024}, likely due to a reduced overlap integral.

The SPDC spectrum from QOM-B (Figure~\ref{fig:SPDC}(d)) shows only a clear narrow peak with a FWHM of 4.6~nm centered at $\sim$1581.9~nm, indicating frequency-degenerate SPDC. The effect of the ED-qBIC resonance in this metasurface is not observable because of a non-optimal pump polarization and a low detection efficiency at both the resonance wavelength and its conjugate. Note that, as discussed earlier, the effect of the in-plane Mie mode is also not observed.

Then, we evaluate the performance of our QOMs by measuring the coincidence rate at various pump powers. For these measurements, we removed the fiber spools.
Figures~\ref{fig:SPDC}(e) and (f) show the real coincidence rates for QOM-A and QOM-B, respectively, calculated by subtracting the accidental coincidences from the total number of coincidences. For QOM-A we present two data sets. The red diamonds represent the real coincidence rate that includes the contribution of both driving resonances -- the ED-qBIC and the in-plane Mie mode.
To isolate the SPDC efficiency due to the ED-qBIC resonance, we subtracted the contribution of the in-plane Mie-driven SPDC in each data point using the measured SPDC spectrum shown in Figure~\ref{fig:SPDC}(d). This isolated ED-qBIC contribution is represented by the blue circles. The data show the characteristic linear dependence of the SPDC rate on the pump power, with dashed lines representing fits of the expected linear relationship.
At an excitation power of 55~mW of excitation power, we observe real coincidence rates of 0.69~$\pm$~0.03~Hz and 2.50~$\pm$~0.11~Hz for the ED-qBIC and in-plane Mie resonances, respectively, in QOM-A, and 7.56~$\pm$~0.15~Hz for the MD-qBIC in QOM-B.
We attribute the difference in coincidence rates between the two qBICs to their distinct $Q$-factors and overlap integrals. While the $Q$-factor sets an upper bound for the conversion efficiency, a suboptimal overlap integral can further reduce this efficiency.
Notably, the rate obtained from the MD-qBIC resonance is 17 times higher than that reported in a similar metasurface made of [001]-GaAs~\cite{Santiago-Cruz_Science_2022}.
In particular, accounting for the pump power and spectral bandwidth, the rate for the MD-qBIC in QOM-B translates to $2.5\times10^{-2}$~Hz/(mW$\cdot$nm), which is an order of magnitude higher than any other SPDC rates previously reported in metasurfaces.~\cite{Santiago-Cruz_NanoLett_2021, Santiago-Cruz_Science_2022, Zhang_SciAdv_2022, MA_NanoLett_2023, Weissflog_Nanophotonics_2024, Ma_arxiv_2024b}.
However, the rate from the ED-qBIC alone is comparable to that of [001]-GaAs metasurface, which we attribute to the differences in $Q$-factors and pump beam diameters. 
In this work, the $Q$-factor of the ED-BIC is 250, and the pump beam diameter 200~$\mu$m, compared to 330 and 140~$\mu$m in the previous work~\cite{Santiago-Cruz_Science_2022}.
The pump beam diameter is crucial because, although the effects may be linear with respect to pump power, the coupling efficiency is affected by the spatial mode structure.
Specifically, a smaller pump area excites fewer spatial modes, thereby enhancing the collection efficiency into the fiber.

\begin{figure}[t!]
\centering
\includegraphics[width=80mm]{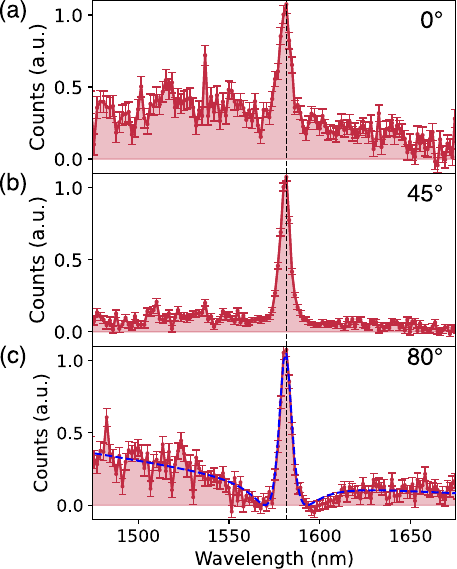}
\caption{Measured SPDC spectra from QOM-A with the polarization analyzer added to the SPDC spectrum measurement setup, where the HWP was rotated to filter various polarization angles: (a) $0^{\circ}$, (b) $45^{\circ}$ and (c) $80^{\circ}$, with the polarization parallel to the horizontal axis denoted as $0^{\circ}$. The blue dashed line shows the fit to the interference of two Lorentzian-shaped resonances, while the black dashed lines indicate double the wavelength of the pump beam.}
\label{fig:Fano}
\end{figure}

Finally, we observe two-photon interference in the spectral domain between biphotons generated by SPDC processes driven by the ED-qBIC mode and the in-plane Mie mode, respectively, in QOM-A. For the demonstration of the two-photon interference, we reinsert 2~km-spools of single-mode fibers into each detection arm for the spectral measurements, along with a HWP and a polarizing beam-splitter (PBS) placed immediately before the NPBS to serve as a polarization analyzer (see Figure~\ref{fig:setup}(b)).

In general, biphotons emitted from different resonance modes exhibit different polarization states\cite{Noh_NanoLett_2024}, and thus, are distinguishable. With the polarization analyzer, we make a projection onto one polarization state and erase any possible distinguishability between the biphotons.
In Figure~\ref{fig:Fano}, we show the measured SPDC spectra from QOM-A with varying filtered polarization angles, with the polarization parallel to the horizontal axis denoted as $0^{\circ}$.
As we change the angle of the analyzer, we observe the expected change in the ratio between the contributions from the qBIC and in-plane Mie resonances. When both contributions are comparable (Figure~\ref{fig:Fano}(c)), we observe a clear Fano contour in the spectrum resulting from the two-photon interference between biphotons emitted by the two different SPDC sources. 
After the analyzer, the experiment cannot distinguish whether the biphotons were generated by the qBIC or by the in-plane Mie mode, and the coincidence measurement exhibits two-photon interference. According to Feynman’s indistinguishability criterion~\cite{Feynman1965}, if the biphotons cannot be distinguished --even in principle--, the individual probability amplitudes should be summed, and the modulus squared should then be calculated to determine the joint detection probability.
The dashed curve in Figure~\ref{fig:Fano}(c) is a fit obtained by taking the modulus square of the sum of the amplitudes of two Lorentzian functions, having as fitting parameters the resonance wavelengths of the qBIC and in-plane Mie modes, their linewidths, and amplitudes, see SI.
The perfect agreement with the experimental data confirms Feynman’s statement and the quantum nature of the interference.
To our knowledge, this is the first demonstration of a quantum two-photon interference enabled by biphoton emission from spatially overlapping yet distinct sources in a metasurface.

To confirm that the Fano-like spectrum arises due to the indistinguishability of the photons, we conducted additional measurements by varying the ratio of contributions between the high-$Q$ qBIC and the low-$Q$ in-plane Mie resonances.
This was done by adjusting the polarization of the pump beam, while making the biphotons distinguishable by taking out the polarization analyzer (see Figure~\ref{fig:NoFano}). 
The SPDC spectra measured in this manner clearly show that the Fano contour, with a distinct dip approaching zero counts, cannot be observed, thereby confirming that the Fano contour seen in Figure~\ref{fig:Fano}(c) is indeed due to the biphoton interference.

In summary, we have demonstrated quantum two-photon interference, a phenomenon crucial for quantum state engineering, in QOMs through biphoton emission from spatially overlapping yet distinct sources, i.e., photonic modes in [110]-GaAs metasurfaces.
This interference was facilitated by the increased SPDC rates, achieved by exploiting the favorable $\chi^{(2)}$ tensor orientation of  [110]-GaAs; furthermore, the quantum interference manifests as a distinctive Fano contour in the spectral analysis when the distinguishability between biphotons from high-$Q$ qBIC and low-$Q$  in-plane Mie resonances is erased by a polarization analyzer.
Beyond the fundamental significance of observing quantum two-photon interference in metasurfaces for the first time, our work establishes [110]-GaAs metasurfaces as a powerful platform for quantum photonics, offering enhanced nonlinear response, fabrication advantages, and unprecedented multifunctionality.

While the demonstrated results are a significant step forward, challenges remain for the practical deployment of metasurface-based SPDC sources~\cite{Santiago-Cruz_NanoLett_2021, Zhang_SciAdv_2022,Santiago-Cruz_Science_2022,Son_Nanoscale_2023,MA_NanoLett_2023,Weissflog_Nanophotonics_2024,Jia_arXiv_2024,Ma_arxiv_2024, Ma_arxiv_2024b}, in real-world quantum technologies. In particular, the issue of unwanted strong photoluminescence (PL),~\cite{Santiago-Cruz_Science_2022,Son_Nanoscale_2023,Santiago-Cruz_OptLett_2021} accompanying SPDC emission in second-order nonlinear zinc-blende materials, pumped above the bandgap, remains a key hurdle for applications such as quantum imaging with undetected photons,~\cite{Lemos_Nature_2014} where no coincidence detection is necessary. To address these limitations, we suggest further exploration of second-order nonlinear zinc-materials materials with higher energy bandgaps, such as AlGaAs, which offer greater transparency in the red wavelength range while preserving comparable second-order nonlinearity. Moreover, materials with even higher energy bandgaps, such as InGaP~\cite{Ma_arxiv_2024b} or GaP, offer potential solutions, with GaP being particularly useful for mid-infrared applications~\cite{Kviatkovsky_SciAdv_2020,Paterova_SciAdv_2020} such as microscopy with undetected photons.

\section*{Data availability}
The data that support the findings of this study are available from the corresponding author on reasonable request.

\section*{Acknowledgements}
This work was supported by the U.S. Department of Energy (DOE), Office of Basic Energy Sciences, Division of Materials Sciences and Engineering and performed, in part, at the Center for Integrated Nanotechnologies, an Office of Science User Facility operated for the U.S. DOE Office of Science.
Sandia National Laboratories is a multi-mission laboratory managed and operated by National Technology and Engineering Solutions of Sandia, LLC., a wholly owned subsidiary of Honeywell International, Inc., for the U.S. DOE’s National Nuclear Security Administration under contract DE-NA0003525.
This paper describes objective technical results and analysis. Any subjective views or opinions that might be expressed in the paper do not necessarily represent the views of the U.S. DOE or the United States Government.
M.V.C is part of the Max Planck School of Photonics supported by BMBF, Max Planck Society and Fraunhofer Society and was funded by the Deutsche Forschungsgemeinschaft (DFG, German Research Foundation), Project 311185701.

\section*{Competing interests}
The authors declare no competing interests.
\clearpage

\pagebreak
\widetext
\begin{center}
\textbf{\large Supplemental Materials: Quantum Pair Generation in Nonlinear Metasurfaces with Mixed and Pure Photon Polarizations}
\end{center}
\setcounter{equation}{0}
\setcounter{figure}{0}
\setcounter{table}{0}
\setcounter{page}{1}
\makeatletter
\renewcommand{\theequation}{S\arabic{equation}}
\renewcommand{\thefigure}{S\arabic{figure}}
\renewcommand{\thetable}{S\arabic{table}}

\section*{I. Nonlinear SHG simulations}
Employing the standard undepleted pump approximation, we followed a two-step process for the simulation: First, we simulate the linear response of the metasurface at the fundamental wavelength, where the pump beam is incident from the air side.
The resulting local field distributions are then used to compute the induced bulk second-order polarization inside the meta-atom. 
Second, this bulk second-order polarization serves as the sole source term for the electromagnetic simulation at the harmonic wavelength, generating the SH field, and we compute the far-field SHG intensity within the effective collection NA. 
Surface nonlinear effects~\cite{Gennaro_ACSPhotonics_2022} were not considered in this simulation. Under optimum conditions, where the pump beam polarization and the metasurface orientation with respect to the crystalline orientations were chosen to maximize the overlap integral between the $\chi^{(2)}$ tensor and the field profiles, the simulated SHG nonlinear conversion factors of QOM-A (ED-qBIC) and QOM-B (MD-qBIC) in [110]-GaAs were 23- and 11-fold higher, respectively, than the corresponding optimum results in [001]-GaAs.

\section*{II. Dimensions of the metasurfaces}
\begin{table}[htbp!]
\centering
\begin{tabular}{|c|c|c|c|c|}
\hline
\textbf{Metasurfaces} & \textbf{\begin{tabular}[c]{@{}c@{}}Cube length,\\ a (nm)\end{tabular}} & \textbf{\begin{tabular}[c]{@{}c@{}}Notch width,\\ w (nm)\end{tabular}} & \textbf{\begin{tabular}[c]{@{}c@{}}Notch length,\\ l (nm)\end{tabular}} & \textbf{Period (nm)} \\ \hline
QOM-A & 376 & 118 & 196 & 795 \\ \hline
QOM-B & 351 & 107 & 175 & 714 \\ \hline
\end{tabular}
\caption{Dimensions of various metasurfaces (QOM-A and QOM-B) considered in this work. The dimensions were carefully chosen such that ED-qBIC and MD-qBIC, respectively, are placed at approximately twice the pump wavelength.} 
\label{tab:dimensions}
\end{table}

\newpage
\section*{III. Metasurface Fabrication}

\begin{figure}[htbp!]
\centering
\includegraphics[width=0.5\textwidth]{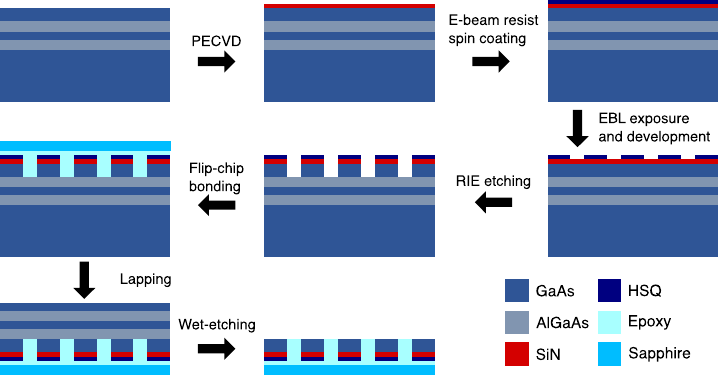}
\caption{Fabrication process for [110]-oriented GaAs metasurface.}
\label{fig:FabProcess}
\end{figure}
To fabricate the QOMs, we utilized a 500~nm-thick [110]-oriented GaAs film grown by molecular beam epitaxy. The GaAs wafer was grown off-orientated by 6$^{\circ}$ towards [111]A to improve surface quality. The metasurfaces comprise square arrays of broken-symmetry meta-atoms (Figure~\ref{fig:Intro}, with detailed dimensions provided in the SI. The fabrication process, illustrated in Figure~\ref{fig:FabProcess}, involved several steps to ensure optimal performance. We began with electron-beam lithography (JEOL, JBX-6300FS) using a negative-tone resist, 6$\%$ hydrogen silsesquioxane (HSQ) solution in MIBK. To address adhesion issues caused by the inherent roughness of the [110]-GaAs surface, we deposited a 7~nm SiN layer by plasma-enhanced chemical vapor deposition (PECVD) before applying the HSQ. After patterning and development (25$\%$ TMAH, 85$^{\circ}$C, 30s), we removed the SiN layer and etched the GaAs using reactive ion etching (RIE) with a gas mixture of BCl$_{3}$, Cl$_{2}$, Ar, and N$_{2}$ (10:10:10:1.5~sccm). The sample was then flip-chip bonded to a sapphire substrate using epoxy (353ND, EPO-TEK), followed by GaAs substrate removal through mechanical lapping and wet-etching with citric acid and phosphoric acid solutions.

\newpage
\section*{IV. Experimental Setups}
\begin{figure}[htbp!]
\centering
\includegraphics[width=0.5\textwidth]{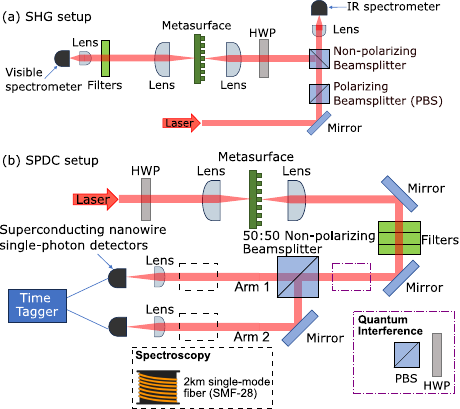}
\caption{(a) Schematics of SHG setup. A pulsed laser (350 fs, 1MHz) pumps the GaAs metasurfaces from the air side. The emitted SHG radiation is filtered off the pump laser using a shortpass filter, and measured with visible spectrometer. The transmission port of the non-polarizing beam splitter (90$\%$ transmission, 10$\%$ reflection) is used to record the pump intensity in an infrared (IR) spectrometer. (b) Schematics of SPDC setup. A continuous-wave laser centered at 788.4 nm pumps the GaAs metasurfaces from the substrate side. The emitted SPDC radiation is filtered off the pump laser using a combination of longpass filters, and then sent to a Hanbury-Brown-Twiss-like interferometer for further analysis and detection. For spectroscopic measurements 2km single-mode fibers are added at the positions indicated with dashed boxes, and for quantum interference measurements additional optical components are further added at the position indicated with dash-dotted box.}
\label{fig:setup}
\end{figure}

\section*{V. Fitting Lorentzian Functions to Experimental Data for Quantum Interference}

To capture the fundamental indistinguishability of biphotons according to Feynman's indistinguishability criterion, we employed the modulus square of the combined amplitudes of two Lorentzian functions, which can be expressed mathematically as follows:

\begin{equation}
    I(\lambda)=\left|\frac{A_{1}\frac{\Gamma_{1}}{2}^{2}}{\pi\left((\lambda-\lambda_{1})^{2}-\frac{\Gamma_{1}}{2}^{2}\right)}-e^{i\phi}\frac{A_{2}\frac{\Gamma_{2}}{2}^{2}}{\pi\left((\lambda-\lambda_{2})^{2}-\frac{\Gamma_{2}}{2}^{2}\right)}\right|^{2},
\end{equation}
where fitting parameters $A_{1,2}$, $\lambda_{1,2}$ and $\Gamma_{1,2}$ are amplitudes, center wavelengths and full-widths at half-maximum of two Lorentzian functions corresponding to the qBIC and the in-plane Mie mode, respectively, and $\phi$ is the phase between them.
$\phi$ from the fit, which is shown as the blue dashed curve in Figure~\ref{fig:Fano}(c), is precisely $\pi$ and other fitting parameters are shown in Table~\ref{tab:fit_parameters}.

\begin{table}[htbp!]
\centering
\begin{tabular}{|c|c|c|c|l}
\cline{1-4}
Modes             & \begin{tabular}[c]{@{}c@{}}Center\\ Wavelength (nm)\end{tabular} & \begin{tabular}[c]{@{}c@{}}Full-width\\ half-maximum (nm)\end{tabular} & Amplitude (a.u.) &  \\ \cline{1-4}
qBIC              & 1581.1 $\pm$ 0.2 & 14.2 $\pm$ 0.9  & 33 $\pm$ 2 &  \\ \cline{1-4}
In-plane Mie mode & 1450 $\pm$ 40 & 400 $\pm$ 60  & 380 $\pm$ 80 &  \\ \cline{1-4}
\end{tabular}
\caption{Fitting parameters for the blue dashed curve in Figure~\ref{fig:Fano}(c).} 
\label{tab:fit_parameters}
\end{table}

\section*{VI. SPDC spectra with varied pump polarization while maintaining photon distinguishability}
\begin{figure}[htbp!]
\centering
\includegraphics[width=80mm]{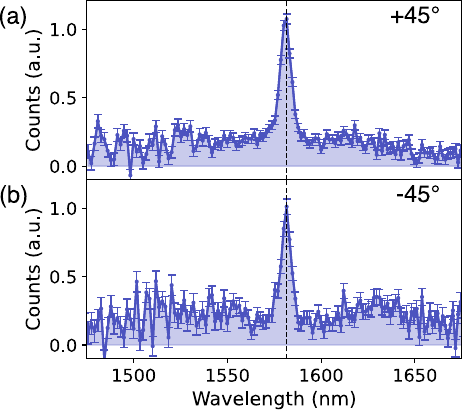}
\caption{(a) Measured SPDC spectra from QOM-A when the linear polarization of the pump beam was rotated 45$^{\circ}$ and (b) -45$^{\circ}$ with respect to that in Figure~\ref{fig:SPDC}(c), where the pump polarization was selected to optimize the nonlinear interaction. Changing the polarization of the pump beam also changes the ratio between the contributions from qBIC and in-plane Mie resonances but still preserves the distinguishability of the photon polarizations. Therefore, unlike in Figure~\ref{fig:Fano}(c), the Fano contour, with a clear dip approaching zero, cannot be observed in the SPDC spectra. Black dashed lines indicate double the wavelength of the pump beam.}
\label{fig:NoFano}
\end{figure}
\clearpage
\newpage

\begin{thebibliography}{46}%
\makeatletter
\providecommand \@ifxundefined [1]{%
 \@ifx{#1\undefined}
}%
\providecommand \@ifnum [1]{%
 \ifnum #1\expandafter \@firstoftwo
 \else \expandafter \@secondoftwo
 \fi
}%
\providecommand \@ifx [1]{%
 \ifx #1\expandafter \@firstoftwo
 \else \expandafter \@secondoftwo
 \fi
}%
\providecommand \natexlab [1]{#1}%
\providecommand \enquote  [1]{``#1''}%
\providecommand \bibnamefont  [1]{#1}%
\providecommand \bibfnamefont [1]{#1}%
\providecommand \citenamefont [1]{#1}%
\providecommand \href@noop [0]{\@secondoftwo}%
\providecommand \href [0]{\begingroup \@sanitize@url \@href}%
\providecommand \@href[1]{\@@startlink{#1}\@@href}%
\providecommand \@@href[1]{\endgroup#1\@@endlink}%
\providecommand \@sanitize@url [0]{\catcode `\\12\catcode `\$12\catcode `\&12\catcode `\#12\catcode `\^12\catcode `\_12\catcode `\%12\relax}%
\providecommand \@@startlink[1]{}%
\providecommand \@@endlink[0]{}%
\providecommand \url  [0]{\begingroup\@sanitize@url \@url }%
\providecommand \@url [1]{\endgroup\@href {#1}{\urlprefix }}%
\providecommand \urlprefix  [0]{URL }%
\providecommand \Eprint [0]{\href }%
\providecommand \doibase [0]{http://dx.doi.org/}%
\providecommand \selectlanguage [0]{\@gobble}%
\providecommand \bibinfo  [0]{\@secondoftwo}%
\providecommand \bibfield  [0]{\@secondoftwo}%
\providecommand \translation [1]{[#1]}%
\providecommand \BibitemOpen [0]{}%
\providecommand \bibitemStop [0]{}%
\providecommand \bibitemNoStop [0]{.\EOS\space}%
\providecommand \EOS [0]{\spacefactor3000\relax}%
\providecommand \BibitemShut  [1]{\csname bibitem#1\endcsname}%
\let\auto@bib@innerbib\@empty
\bibitem [{\citenamefont {Stobi{\'n}ska}\ \emph {et~al.}(2019)\citenamefont {Stobi{\'n}ska}, \citenamefont {Buraczewski}, \citenamefont {Moore}, \citenamefont {Clements}, \citenamefont {Renema}, \citenamefont {Nam}, \citenamefont {Gerrits}, \citenamefont {Lita}, \citenamefont {Kolthammer}, \citenamefont {Eckstein},\ and\ \citenamefont {Walmsley}}]{Stobinska_SciAdv_2019}%
  \BibitemOpen
  \bibfield  {author} {\bibinfo {author} {\bibfnamefont {M.}~\bibnamefont {Stobi{\'n}ska}}, \bibinfo {author} {\bibfnamefont {A.}~\bibnamefont {Buraczewski}}, \bibinfo {author} {\bibfnamefont {M.}~\bibnamefont {Moore}}, \bibinfo {author} {\bibfnamefont {W.~R.}\ \bibnamefont {Clements}}, \bibinfo {author} {\bibfnamefont {J.~J.}\ \bibnamefont {Renema}}, \bibinfo {author} {\bibfnamefont {S.~W.}\ \bibnamefont {Nam}}, \bibinfo {author} {\bibfnamefont {T.}~\bibnamefont {Gerrits}}, \bibinfo {author} {\bibfnamefont {A.}~\bibnamefont {Lita}}, \bibinfo {author} {\bibfnamefont {W.~S.}\ \bibnamefont {Kolthammer}}, \bibinfo {author} {\bibfnamefont {A.}~\bibnamefont {Eckstein}}, \ and\ \bibinfo {author} {\bibfnamefont {I.~A.}\ \bibnamefont {Walmsley}},\ }\href@noop {} {\bibfield  {journal} {\bibinfo  {journal} {Sci. Adv.}\ }\textbf {\bibinfo {volume} {5}},\ \bibinfo {pages} {eaau9674} (\bibinfo {year} {2019})}\BibitemShut {NoStop}%
\bibitem [{\citenamefont {Qiang}\ \emph {et~al.}(2018)\citenamefont {Qiang}, \citenamefont {Zhou}, \citenamefont {Wang}, \citenamefont {Wilkes}, \citenamefont {Loke}, \citenamefont {O'Gara}, \citenamefont {Kling}, \citenamefont {Marshall}, \citenamefont {Santagati}, \citenamefont {Ralph}, \citenamefont {Wang}, \citenamefont {O'Brien}, \citenamefont {Thompson},\ and\ \citenamefont {Matthews}}]{Qiang_NatPhoton_2018}%
  \BibitemOpen
  \bibfield  {author} {\bibinfo {author} {\bibfnamefont {X.}~\bibnamefont {Qiang}}, \bibinfo {author} {\bibfnamefont {X.}~\bibnamefont {Zhou}}, \bibinfo {author} {\bibfnamefont {J.}~\bibnamefont {Wang}}, \bibinfo {author} {\bibfnamefont {C.~M.}\ \bibnamefont {Wilkes}}, \bibinfo {author} {\bibfnamefont {T.}~\bibnamefont {Loke}}, \bibinfo {author} {\bibfnamefont {S.}~\bibnamefont {O'Gara}}, \bibinfo {author} {\bibfnamefont {L.}~\bibnamefont {Kling}}, \bibinfo {author} {\bibfnamefont {G.~D.}\ \bibnamefont {Marshall}}, \bibinfo {author} {\bibfnamefont {R.}~\bibnamefont {Santagati}}, \bibinfo {author} {\bibfnamefont {T.~C.}\ \bibnamefont {Ralph}}, \bibinfo {author} {\bibfnamefont {J.~B.}\ \bibnamefont {Wang}}, \bibinfo {author} {\bibfnamefont {J.~L.}\ \bibnamefont {O'Brien}}, \bibinfo {author} {\bibfnamefont {M.~G.}\ \bibnamefont {Thompson}}, \ and\ \bibinfo {author} {\bibfnamefont {J.~C.~F.}\ \bibnamefont {Matthews}},\ }\href@noop {} {\bibfield  {journal} {\bibinfo  {journal} {Nat. Photonics}\ }\textbf {\bibinfo
  {volume} {12}},\ \bibinfo {pages} {534} (\bibinfo {year} {2018})}\BibitemShut {NoStop}%
\bibitem [{\citenamefont {Knill}\ \emph {et~al.}(2001)\citenamefont {Knill}, \citenamefont {Laflamme},\ and\ \citenamefont {Milburn}}]{Knill_Nature_2001}%
  \BibitemOpen
  \bibfield  {author} {\bibinfo {author} {\bibfnamefont {E.}~\bibnamefont {Knill}}, \bibinfo {author} {\bibfnamefont {R.}~\bibnamefont {Laflamme}}, \ and\ \bibinfo {author} {\bibfnamefont {G.~J.}\ \bibnamefont {Milburn}},\ }\href@noop {} {\bibfield  {journal} {\bibinfo  {journal} {Nature}\ }\textbf {\bibinfo {volume} {409}},\ \bibinfo {pages} {46} (\bibinfo {year} {2001})}\BibitemShut {NoStop}%
\bibitem [{\citenamefont {Spring}\ \emph {et~al.}(2013)\citenamefont {Spring}, \citenamefont {Metcalf}, \citenamefont {Humphreys}, \citenamefont {Kolthammer}, \citenamefont {Jin}, \citenamefont {Barbieri}, \citenamefont {Datta}, \citenamefont {Thomas-Peter}, \citenamefont {Langford}, \citenamefont {Kundys}, \citenamefont {Gates}, \citenamefont {Smith}, \citenamefont {Smith},\ and\ \citenamefont {Walmsley}}]{Spring_Science_2013}%
  \BibitemOpen
  \bibfield  {author} {\bibinfo {author} {\bibfnamefont {J.~B.}\ \bibnamefont {Spring}}, \bibinfo {author} {\bibfnamefont {B.~J.}\ \bibnamefont {Metcalf}}, \bibinfo {author} {\bibfnamefont {P.~C.}\ \bibnamefont {Humphreys}}, \bibinfo {author} {\bibfnamefont {W.~S.}\ \bibnamefont {Kolthammer}}, \bibinfo {author} {\bibfnamefont {X.-M.}\ \bibnamefont {Jin}}, \bibinfo {author} {\bibfnamefont {M.}~\bibnamefont {Barbieri}}, \bibinfo {author} {\bibfnamefont {A.}~\bibnamefont {Datta}}, \bibinfo {author} {\bibfnamefont {N.}~\bibnamefont {Thomas-Peter}}, \bibinfo {author} {\bibfnamefont {N.~K.}\ \bibnamefont {Langford}}, \bibinfo {author} {\bibfnamefont {D.}~\bibnamefont {Kundys}}, \bibinfo {author} {\bibfnamefont {J.~C.}\ \bibnamefont {Gates}}, \bibinfo {author} {\bibfnamefont {B.~J.}\ \bibnamefont {Smith}}, \bibinfo {author} {\bibfnamefont {P.~G.~R.}\ \bibnamefont {Smith}}, \ and\ \bibinfo {author} {\bibfnamefont {I.~A.}\ \bibnamefont {Walmsley}},\ }\href@noop {} {\bibfield  {journal} {\bibinfo  {journal} {Science}\
  }\textbf {\bibinfo {volume} {339}},\ \bibinfo {pages} {798} (\bibinfo {year} {2013})}\BibitemShut {NoStop}%
\bibitem [{\citenamefont {Tillmann}\ \emph {et~al.}(2013)\citenamefont {Tillmann}, \citenamefont {Daki{\'c}}, \citenamefont {Heilmann}, \citenamefont {Nolte}, \citenamefont {Szameit},\ and\ \citenamefont {Walther}}]{Tillmann_NatPhoton_2013}%
  \BibitemOpen
  \bibfield  {author} {\bibinfo {author} {\bibfnamefont {M.}~\bibnamefont {Tillmann}}, \bibinfo {author} {\bibfnamefont {B.}~\bibnamefont {Daki{\'c}}}, \bibinfo {author} {\bibfnamefont {R.}~\bibnamefont {Heilmann}}, \bibinfo {author} {\bibfnamefont {S.}~\bibnamefont {Nolte}}, \bibinfo {author} {\bibfnamefont {A.}~\bibnamefont {Szameit}}, \ and\ \bibinfo {author} {\bibfnamefont {P.}~\bibnamefont {Walther}},\ }\href@noop {} {\bibfield  {journal} {\bibinfo  {journal} {Nat. Photonics}\ }\textbf {\bibinfo {volume} {7}},\ \bibinfo {pages} {540} (\bibinfo {year} {2013})}\BibitemShut {NoStop}%
\bibitem [{\citenamefont {Bao}\ \emph {et~al.}(2023)\citenamefont {Bao}, \citenamefont {Fu}, \citenamefont {Pramanik}, \citenamefont {Mao}, \citenamefont {Chi}, \citenamefont {Cao}, \citenamefont {Zhai}, \citenamefont {Mao}, \citenamefont {Dai}, \citenamefont {Chen}, \citenamefont {Jia}, \citenamefont {Zhao}, \citenamefont {Zheng}, \citenamefont {Tang}, \citenamefont {Li}, \citenamefont {Luo}, \citenamefont {Wang}, \citenamefont {Yang}, \citenamefont {Peng}, \citenamefont {Liu}, \citenamefont {Dai}, \citenamefont {He}, \citenamefont {Muthali}, \citenamefont {Oxenl{\o}we}, \citenamefont {Vigliar}, \citenamefont {Paesani}, \citenamefont {Hou}, \citenamefont {Santagati}, \citenamefont {Silverstone}, \citenamefont {Laing}, \citenamefont {Thompson}, \citenamefont {O'Brien}, \citenamefont {Ding}, \citenamefont {Gong},\ and\ \citenamefont {Wang}}]{Bao_NatPhoton_2023}%
  \BibitemOpen
  \bibfield  {author} {\bibinfo {author} {\bibfnamefont {J.}~\bibnamefont {Bao}}, \bibinfo {author} {\bibfnamefont {Z.}~\bibnamefont {Fu}}, \bibinfo {author} {\bibfnamefont {T.}~\bibnamefont {Pramanik}}, \bibinfo {author} {\bibfnamefont {J.}~\bibnamefont {Mao}}, \bibinfo {author} {\bibfnamefont {Y.}~\bibnamefont {Chi}}, \bibinfo {author} {\bibfnamefont {Y.}~\bibnamefont {Cao}}, \bibinfo {author} {\bibfnamefont {C.}~\bibnamefont {Zhai}}, \bibinfo {author} {\bibfnamefont {Y.}~\bibnamefont {Mao}}, \bibinfo {author} {\bibfnamefont {T.}~\bibnamefont {Dai}}, \bibinfo {author} {\bibfnamefont {X.}~\bibnamefont {Chen}}, \bibinfo {author} {\bibfnamefont {X.}~\bibnamefont {Jia}}, \bibinfo {author} {\bibfnamefont {L.}~\bibnamefont {Zhao}}, \bibinfo {author} {\bibfnamefont {Y.}~\bibnamefont {Zheng}}, \bibinfo {author} {\bibfnamefont {B.}~\bibnamefont {Tang}}, \bibinfo {author} {\bibfnamefont {Z.}~\bibnamefont {Li}}, \bibinfo {author} {\bibfnamefont {J.}~\bibnamefont {Luo}}, \bibinfo {author} {\bibfnamefont
  {W.}~\bibnamefont {Wang}}, \bibinfo {author} {\bibfnamefont {Y.}~\bibnamefont {Yang}}, \bibinfo {author} {\bibfnamefont {Y.}~\bibnamefont {Peng}}, \bibinfo {author} {\bibfnamefont {D.}~\bibnamefont {Liu}}, \bibinfo {author} {\bibfnamefont {D.}~\bibnamefont {Dai}}, \bibinfo {author} {\bibfnamefont {Q.}~\bibnamefont {He}}, \bibinfo {author} {\bibfnamefont {A.~L.}\ \bibnamefont {Muthali}}, \bibinfo {author} {\bibfnamefont {L.~K.}\ \bibnamefont {Oxenl{\o}we}}, \bibinfo {author} {\bibfnamefont {C.}~\bibnamefont {Vigliar}}, \bibinfo {author} {\bibfnamefont {S.}~\bibnamefont {Paesani}}, \bibinfo {author} {\bibfnamefont {H.}~\bibnamefont {Hou}}, \bibinfo {author} {\bibfnamefont {R.}~\bibnamefont {Santagati}}, \bibinfo {author} {\bibfnamefont {J.~W.}\ \bibnamefont {Silverstone}}, \bibinfo {author} {\bibfnamefont {A.}~\bibnamefont {Laing}}, \bibinfo {author} {\bibfnamefont {M.~G.}\ \bibnamefont {Thompson}}, \bibinfo {author} {\bibfnamefont {J.~L.}\ \bibnamefont {O'Brien}}, \bibinfo {author} {\bibfnamefont
  {Y.}~\bibnamefont {Ding}}, \bibinfo {author} {\bibfnamefont {Q.}~\bibnamefont {Gong}}, \ and\ \bibinfo {author} {\bibfnamefont {J.}~\bibnamefont {Wang}},\ }\href@noop {} {\bibfield  {journal} {\bibinfo  {journal} {Nat. Photonics}\ }\textbf {\bibinfo {volume} {17}},\ \bibinfo {pages} {573} (\bibinfo {year} {2023})}\BibitemShut {NoStop}%
\bibitem [{\citenamefont {Lemos}\ \emph {et~al.}(2014)\citenamefont {Lemos}, \citenamefont {Borish}, \citenamefont {Cole}, \citenamefont {Ramelow}, \citenamefont {Lapkiewicz},\ and\ \citenamefont {Zeilinger}}]{Lemos_Nature_2014}%
  \BibitemOpen
  \bibfield  {author} {\bibinfo {author} {\bibfnamefont {G.~B.}\ \bibnamefont {Lemos}}, \bibinfo {author} {\bibfnamefont {V.}~\bibnamefont {Borish}}, \bibinfo {author} {\bibfnamefont {G.~D.}\ \bibnamefont {Cole}}, \bibinfo {author} {\bibfnamefont {S.}~\bibnamefont {Ramelow}}, \bibinfo {author} {\bibfnamefont {R.}~\bibnamefont {Lapkiewicz}}, \ and\ \bibinfo {author} {\bibfnamefont {A.}~\bibnamefont {Zeilinger}},\ }\href@noop {} {\bibfield  {journal} {\bibinfo  {journal} {Nature}\ }\textbf {\bibinfo {volume} {512}},\ \bibinfo {pages} {409} (\bibinfo {year} {2014})}\BibitemShut {NoStop}%
\bibitem [{\citenamefont {Paterova}\ \emph {et~al.}(2020)\citenamefont {Paterova}, \citenamefont {Maniam}, \citenamefont {Yang}, \citenamefont {Grenci},\ and\ \citenamefont {Krivitsky}}]{Paterova_SciAdv_2020}%
  \BibitemOpen
  \bibfield  {author} {\bibinfo {author} {\bibfnamefont {A.~V.}\ \bibnamefont {Paterova}}, \bibinfo {author} {\bibfnamefont {S.~M.}\ \bibnamefont {Maniam}}, \bibinfo {author} {\bibfnamefont {H.}~\bibnamefont {Yang}}, \bibinfo {author} {\bibfnamefont {G.}~\bibnamefont {Grenci}}, \ and\ \bibinfo {author} {\bibfnamefont {L.~A.}\ \bibnamefont {Krivitsky}},\ }\href@noop {} {\bibfield  {journal} {\bibinfo  {journal} {Sci. Adv.}\ }\textbf {\bibinfo {volume} {6}},\ \bibinfo {pages} {eabd0460} (\bibinfo {year} {2020})}\BibitemShut {NoStop}%
\bibitem [{\citenamefont {Kviatkovsky}\ \emph {et~al.}(2020)\citenamefont {Kviatkovsky}, \citenamefont {Chrzanowski}, \citenamefont {Avery}, \citenamefont {Bartolomaeus},\ and\ \citenamefont {Ramelow}}]{Kviatkovsky_SciAdv_2020}%
  \BibitemOpen
  \bibfield  {author} {\bibinfo {author} {\bibfnamefont {I.}~\bibnamefont {Kviatkovsky}}, \bibinfo {author} {\bibfnamefont {H.~M.}\ \bibnamefont {Chrzanowski}}, \bibinfo {author} {\bibfnamefont {E.~G.}\ \bibnamefont {Avery}}, \bibinfo {author} {\bibfnamefont {H.}~\bibnamefont {Bartolomaeus}}, \ and\ \bibinfo {author} {\bibfnamefont {S.}~\bibnamefont {Ramelow}},\ }\href@noop {} {\bibfield  {journal} {\bibinfo  {journal} {Sci. Adv.}\ }\textbf {\bibinfo {volume} {6}},\ \bibinfo {pages} {eabd0264} (\bibinfo {year} {2020})}\BibitemShut {NoStop}%
\bibitem [{\citenamefont {Clark}\ \emph {et~al.}(2021)\citenamefont {Clark}, \citenamefont {Chekhova}, \citenamefont {Matthews}, \citenamefont {Rarity},\ and\ \citenamefont {Oulton}}]{Clark_ApplPhysLett_2021}%
  \BibitemOpen
  \bibfield  {author} {\bibinfo {author} {\bibfnamefont {A.~S.}\ \bibnamefont {Clark}}, \bibinfo {author} {\bibfnamefont {M.}~\bibnamefont {Chekhova}}, \bibinfo {author} {\bibfnamefont {J.~C.~F.}\ \bibnamefont {Matthews}}, \bibinfo {author} {\bibfnamefont {J.~G.}\ \bibnamefont {Rarity}}, \ and\ \bibinfo {author} {\bibfnamefont {R.~F.}\ \bibnamefont {Oulton}},\ }\href@noop {} {\bibfield  {journal} {\bibinfo  {journal} {Appl. Phys. Lett.}\ }\textbf {\bibinfo {volume} {118}},\ \bibinfo {pages} {060401} (\bibinfo {year} {2021})}\BibitemShut {NoStop}%
\bibitem [{\citenamefont {Defienne}\ \emph {et~al.}(2024)\citenamefont {Defienne}, \citenamefont {Bowen}, \citenamefont {Chekhova}, \citenamefont {Lemos}, \citenamefont {Oron}, \citenamefont {Ramelow}, \citenamefont {Treps},\ and\ \citenamefont {Faccio}}]{Defienne_NatPhoton_2024}%
  \BibitemOpen
  \bibfield  {author} {\bibinfo {author} {\bibfnamefont {H.}~\bibnamefont {Defienne}}, \bibinfo {author} {\bibfnamefont {W.~P.}\ \bibnamefont {Bowen}}, \bibinfo {author} {\bibfnamefont {M.}~\bibnamefont {Chekhova}}, \bibinfo {author} {\bibfnamefont {G.~B.}\ \bibnamefont {Lemos}}, \bibinfo {author} {\bibfnamefont {D.}~\bibnamefont {Oron}}, \bibinfo {author} {\bibfnamefont {S.}~\bibnamefont {Ramelow}}, \bibinfo {author} {\bibfnamefont {N.}~\bibnamefont {Treps}}, \ and\ \bibinfo {author} {\bibfnamefont {D.}~\bibnamefont {Faccio}},\ }\href@noop {} {\bibfield  {journal} {\bibinfo  {journal} {Nat. Photonics}\ }\textbf {\bibinfo {volume} {18}},\ \bibinfo {pages} {1024} (\bibinfo {year} {2024})}\BibitemShut {NoStop}%
\bibitem [{\citenamefont {Mandel}(1999)}]{Mandel_RevModPhys_1999}%
  \BibitemOpen
  \bibfield  {author} {\bibinfo {author} {\bibfnamefont {L.}~\bibnamefont {Mandel}},\ }\href {\doibase 10.1103/RevModPhys.71.S274} {\bibfield  {journal} {\bibinfo  {journal} {Rev. Mod. Phys.}\ }\textbf {\bibinfo {volume} {71}},\ \bibinfo {pages} {S274} (\bibinfo {year} {1999})}\BibitemShut {NoStop}%
\bibitem [{\citenamefont {Feynman}\ \emph {et~al.}(1965)\citenamefont {Feynman}, \citenamefont {Leighton},\ and\ \citenamefont {Sands}}]{Feynman1965}%
  \BibitemOpen
  \bibfield  {author} {\bibinfo {author} {\bibfnamefont {R.~P.}\ \bibnamefont {Feynman}}, \bibinfo {author} {\bibfnamefont {R.~B.}\ \bibnamefont {Leighton}}, \ and\ \bibinfo {author} {\bibfnamefont {M.}~\bibnamefont {Sands}},\ }\href@noop {} {\emph {\bibinfo {title} {The Feynman Lectures on Physics: Volume 3: Quantum Mechanics}}}\ (\bibinfo  {publisher} {Addison-Wesley},\ \bibinfo {address} {Reading, MA},\ \bibinfo {year} {1965})\BibitemShut {NoStop}%
\bibitem [{\citenamefont {Kwiat}\ \emph {et~al.}(1999)\citenamefont {Kwiat}, \citenamefont {Waks}, \citenamefont {White}, \citenamefont {Appelbaum},\ and\ \citenamefont {Eberhard}}]{Kwiat_PhysRevA_1999}%
  \BibitemOpen
  \bibfield  {author} {\bibinfo {author} {\bibfnamefont {P.~G.}\ \bibnamefont {Kwiat}}, \bibinfo {author} {\bibfnamefont {E.}~\bibnamefont {Waks}}, \bibinfo {author} {\bibfnamefont {A.~G.}\ \bibnamefont {White}}, \bibinfo {author} {\bibfnamefont {I.}~\bibnamefont {Appelbaum}}, \ and\ \bibinfo {author} {\bibfnamefont {P.~H.}\ \bibnamefont {Eberhard}},\ }\href {\doibase 10.1103/PhysRevA.60.R773} {\bibfield  {journal} {\bibinfo  {journal} {Phys. Rev. A}\ }\textbf {\bibinfo {volume} {60}},\ \bibinfo {pages} {R773} (\bibinfo {year} {1999})}\BibitemShut {NoStop}%
\bibitem [{\citenamefont {Burlakov}\ \emph {et~al.}(2001)\citenamefont {Burlakov}, \citenamefont {Chekhova}, \citenamefont {Karabutova},\ and\ \citenamefont {Kulik}}]{Burlakov_PRA_2001}%
  \BibitemOpen
  \bibfield  {author} {\bibinfo {author} {\bibfnamefont {A.~V.}\ \bibnamefont {Burlakov}}, \bibinfo {author} {\bibfnamefont {M.~V.}\ \bibnamefont {Chekhova}}, \bibinfo {author} {\bibfnamefont {O.~A.}\ \bibnamefont {Karabutova}}, \ and\ \bibinfo {author} {\bibfnamefont {S.~P.}\ \bibnamefont {Kulik}},\ }\href {\doibase 10.1103/PhysRevA.64.041803} {\bibfield  {journal} {\bibinfo  {journal} {Phys. Rev. A}\ }\textbf {\bibinfo {volume} {64}},\ \bibinfo {pages} {041803} (\bibinfo {year} {2001})}\BibitemShut {NoStop}%
\bibitem [{\citenamefont {Brida}\ \emph {et~al.}(2006)\citenamefont {Brida}, \citenamefont {Chekhova}, \citenamefont {Genovese}, \citenamefont {Gramegna},\ and\ \citenamefont {Krivitsky}}]{Brida_PRL_2006}%
  \BibitemOpen
  \bibfield  {author} {\bibinfo {author} {\bibfnamefont {G.}~\bibnamefont {Brida}}, \bibinfo {author} {\bibfnamefont {M.~V.}\ \bibnamefont {Chekhova}}, \bibinfo {author} {\bibfnamefont {M.}~\bibnamefont {Genovese}}, \bibinfo {author} {\bibfnamefont {M.}~\bibnamefont {Gramegna}}, \ and\ \bibinfo {author} {\bibfnamefont {L.~A.}\ \bibnamefont {Krivitsky}},\ }\href {\doibase 10.1103/PhysRevLett.96.143601} {\bibfield  {journal} {\bibinfo  {journal} {Phys. Rev. Lett.}\ }\textbf {\bibinfo {volume} {96}},\ \bibinfo {pages} {143601} (\bibinfo {year} {2006})}\BibitemShut {NoStop}%
\bibitem [{\citenamefont {Chekhova}\ and\ \citenamefont {Ou}(2016)}]{Chekhova_AdvOptPhoton_2016}%
  \BibitemOpen
  \bibfield  {author} {\bibinfo {author} {\bibfnamefont {M.~V.}\ \bibnamefont {Chekhova}}\ and\ \bibinfo {author} {\bibfnamefont {Z.~Y.}\ \bibnamefont {Ou}},\ }\href@noop {} {\bibfield  {journal} {\bibinfo  {journal} {Adv. Opt. Photonics}\ }\textbf {\bibinfo {volume} {8}},\ \bibinfo {pages} {104} (\bibinfo {year} {2016})}\BibitemShut {NoStop}%
\bibitem [{\citenamefont {Chen}\ \emph {et~al.}(2016)\citenamefont {Chen}, \citenamefont {Taylor},\ and\ \citenamefont {Yu}}]{Chen_RepProgPhys_2016}%
  \BibitemOpen
  \bibfield  {author} {\bibinfo {author} {\bibfnamefont {H.-T.}\ \bibnamefont {Chen}}, \bibinfo {author} {\bibfnamefont {A.~J.}\ \bibnamefont {Taylor}}, \ and\ \bibinfo {author} {\bibfnamefont {N.}~\bibnamefont {Yu}},\ }\href {\doibase 10.1088/0034-4885/79/7/076401} {\bibfield  {journal} {\bibinfo  {journal} {Reports on Progress in Physics}\ }\textbf {\bibinfo {volume} {79}},\ \bibinfo {pages} {076401} (\bibinfo {year} {2016})}\BibitemShut {NoStop}%
\bibitem [{\citenamefont {Li}\ \emph {et~al.}(2017)\citenamefont {Li}, \citenamefont {Zhang},\ and\ \citenamefont {Zentgraf}}]{Li_NatRevMats_2017}%
  \BibitemOpen
  \bibfield  {author} {\bibinfo {author} {\bibfnamefont {G.}~\bibnamefont {Li}}, \bibinfo {author} {\bibfnamefont {S.}~\bibnamefont {Zhang}}, \ and\ \bibinfo {author} {\bibfnamefont {T.}~\bibnamefont {Zentgraf}},\ }\href {\doibase 10.1038/natrevmats.2017.10} {\bibfield  {journal} {\bibinfo  {journal} {Nature Reviews Materials}\ }\textbf {\bibinfo {volume} {2}},\ \bibinfo {pages} {17010} (\bibinfo {year} {2017})}\BibitemShut {NoStop}%
\bibitem [{\citenamefont {Krasnok}\ \emph {et~al.}(2018)\citenamefont {Krasnok}, \citenamefont {Tymchenko},\ and\ \citenamefont {Al{\`u}}}]{Krasnok_MaterToday_2018}%
  \BibitemOpen
  \bibfield  {author} {\bibinfo {author} {\bibfnamefont {A.}~\bibnamefont {Krasnok}}, \bibinfo {author} {\bibfnamefont {M.}~\bibnamefont {Tymchenko}}, \ and\ \bibinfo {author} {\bibfnamefont {A.}~\bibnamefont {Al{\`u}}},\ }\href@noop {} {\bibfield  {journal} {\bibinfo  {journal} {Mater. Today (Kidlington)}\ }\textbf {\bibinfo {volume} {21}},\ \bibinfo {pages} {8} (\bibinfo {year} {2018})}\BibitemShut {NoStop}%
\bibitem [{\citenamefont {Sain}\ \emph {et~al.}(2019)\citenamefont {Sain}, \citenamefont {Meier},\ and\ \citenamefont {Zentgraf}}]{Sain_AdvPhoton_2019}%
  \BibitemOpen
  \bibfield  {author} {\bibinfo {author} {\bibfnamefont {B.}~\bibnamefont {Sain}}, \bibinfo {author} {\bibfnamefont {C.}~\bibnamefont {Meier}}, \ and\ \bibinfo {author} {\bibfnamefont {T.}~\bibnamefont {Zentgraf}},\ }\href {\doibase 10.1117/1.AP.1.2.024002} {\bibfield  {journal} {\bibinfo  {journal} {Advanced Photonics}\ }\textbf {\bibinfo {volume} {1}},\ \bibinfo {pages} {024002} (\bibinfo {year} {2019})}\BibitemShut {NoStop}%
\bibitem [{\citenamefont {Vabishchevich}\ and\ \citenamefont {Kivshar}(2023)}]{Vabishchevich_PhotonRes_2023}%
  \BibitemOpen
  \bibfield  {author} {\bibinfo {author} {\bibfnamefont {P.}~\bibnamefont {Vabishchevich}}\ and\ \bibinfo {author} {\bibfnamefont {Y.}~\bibnamefont {Kivshar}},\ }\href {\doibase 10.1364/PRJ.474387} {\bibfield  {journal} {\bibinfo  {journal} {Photonics Research}\ }\textbf {\bibinfo {volume} {11}},\ \bibinfo {pages} {B50} (\bibinfo {year} {2023})}\BibitemShut {NoStop}%
\bibitem [{\citenamefont {Santiago-Cruz}\ \emph {et~al.}(2021{\natexlab{a}})\citenamefont {Santiago-Cruz}, \citenamefont {Fedotova}, \citenamefont {Sultanov}, \citenamefont {Weissflog}, \citenamefont {Arslan}, \citenamefont {Younesi}, \citenamefont {Pertsch}, \citenamefont {Staude}, \citenamefont {Setzpfandt},\ and\ \citenamefont {Chekhova}}]{Santiago-Cruz_NanoLett_2021}%
  \BibitemOpen
  \bibfield  {author} {\bibinfo {author} {\bibfnamefont {T.}~\bibnamefont {Santiago-Cruz}}, \bibinfo {author} {\bibfnamefont {A.}~\bibnamefont {Fedotova}}, \bibinfo {author} {\bibfnamefont {V.}~\bibnamefont {Sultanov}}, \bibinfo {author} {\bibfnamefont {M.~A.}\ \bibnamefont {Weissflog}}, \bibinfo {author} {\bibfnamefont {D.}~\bibnamefont {Arslan}}, \bibinfo {author} {\bibfnamefont {M.}~\bibnamefont {Younesi}}, \bibinfo {author} {\bibfnamefont {T.}~\bibnamefont {Pertsch}}, \bibinfo {author} {\bibfnamefont {I.}~\bibnamefont {Staude}}, \bibinfo {author} {\bibfnamefont {F.}~\bibnamefont {Setzpfandt}}, \ and\ \bibinfo {author} {\bibfnamefont {M.~V.}\ \bibnamefont {Chekhova}},\ }\href {\doibase 10.1021/acs.nanolett.1c01125} {\bibfield  {journal} {\bibinfo  {journal} {Nano Letters}\ }\textbf {\bibinfo {volume} {21}},\ \bibinfo {pages} {4423–4429} (\bibinfo {year} {2021}{\natexlab{a}})}\BibitemShut {NoStop}%
\bibitem [{\citenamefont {Santiago-Cruz}\ \emph {et~al.}(2022)\citenamefont {Santiago-Cruz}, \citenamefont {Gennaro}, \citenamefont {Mitrofanov}, \citenamefont {Addamane}, \citenamefont {Reno}, \citenamefont {Brener},\ and\ \citenamefont {Chekhova}}]{Santiago-Cruz_Science_2022}%
  \BibitemOpen
  \bibfield  {author} {\bibinfo {author} {\bibfnamefont {T.}~\bibnamefont {Santiago-Cruz}}, \bibinfo {author} {\bibfnamefont {S.~D.}\ \bibnamefont {Gennaro}}, \bibinfo {author} {\bibfnamefont {O.}~\bibnamefont {Mitrofanov}}, \bibinfo {author} {\bibfnamefont {S.}~\bibnamefont {Addamane}}, \bibinfo {author} {\bibfnamefont {J.}~\bibnamefont {Reno}}, \bibinfo {author} {\bibfnamefont {I.}~\bibnamefont {Brener}}, \ and\ \bibinfo {author} {\bibfnamefont {M.~V.}\ \bibnamefont {Chekhova}},\ }\href {\doibase 10.1126/science.abq8684} {\bibfield  {journal} {\bibinfo  {journal} {Science}\ }\textbf {\bibinfo {volume} {377}},\ \bibinfo {pages} {991–995} (\bibinfo {year} {2022})}\BibitemShut {NoStop}%
\bibitem [{\citenamefont {Zhang}\ \emph {et~al.}(2022)\citenamefont {Zhang}, \citenamefont {Ma}, \citenamefont {Parry}, \citenamefont {Cai}, \citenamefont {Camacho-Morales}, \citenamefont {Xu}, \citenamefont {Neshev},\ and\ \citenamefont {Sukhorukov}}]{Zhang_SciAdv_2022}%
  \BibitemOpen
  \bibfield  {author} {\bibinfo {author} {\bibfnamefont {J.}~\bibnamefont {Zhang}}, \bibinfo {author} {\bibfnamefont {J.}~\bibnamefont {Ma}}, \bibinfo {author} {\bibfnamefont {M.}~\bibnamefont {Parry}}, \bibinfo {author} {\bibfnamefont {M.}~\bibnamefont {Cai}}, \bibinfo {author} {\bibfnamefont {R.}~\bibnamefont {Camacho-Morales}}, \bibinfo {author} {\bibfnamefont {L.}~\bibnamefont {Xu}}, \bibinfo {author} {\bibfnamefont {D.~N.}\ \bibnamefont {Neshev}}, \ and\ \bibinfo {author} {\bibfnamefont {A.~A.}\ \bibnamefont {Sukhorukov}},\ }\href {\doibase 10.1126/sciadv.abq4240} {\bibfield  {journal} {\bibinfo  {journal} {Science Advances}\ }\textbf {\bibinfo {volume} {8}},\ \bibinfo {pages} {eabq4240} (\bibinfo {year} {2022})}\BibitemShut {NoStop}%
\bibitem [{\citenamefont {Son}\ \emph {et~al.}(2023)\citenamefont {Son}, \citenamefont {Sultanov}, \citenamefont {Santiago-Cruz}, \citenamefont {Anthur}, \citenamefont {Zhang}, \citenamefont {Paniagua-Dominguez}, \citenamefont {Krivitsky}, \citenamefont {Kuznetsov},\ and\ \citenamefont {Chekhova}}]{Son_Nanoscale_2023}%
  \BibitemOpen
  \bibfield  {author} {\bibinfo {author} {\bibfnamefont {C.}~\bibnamefont {Son}}, \bibinfo {author} {\bibfnamefont {V.}~\bibnamefont {Sultanov}}, \bibinfo {author} {\bibfnamefont {T.}~\bibnamefont {Santiago-Cruz}}, \bibinfo {author} {\bibfnamefont {A.~P.}\ \bibnamefont {Anthur}}, \bibinfo {author} {\bibfnamefont {H.}~\bibnamefont {Zhang}}, \bibinfo {author} {\bibfnamefont {R.}~\bibnamefont {Paniagua-Dominguez}}, \bibinfo {author} {\bibfnamefont {L.}~\bibnamefont {Krivitsky}}, \bibinfo {author} {\bibfnamefont {A.~I.}\ \bibnamefont {Kuznetsov}}, \ and\ \bibinfo {author} {\bibfnamefont {M.~V.}\ \bibnamefont {Chekhova}},\ }\href {\doibase 10.1039/D2NR05499J} {\bibfield  {journal} {\bibinfo  {journal} {Nanoscale}\ }\textbf {\bibinfo {volume} {15}},\ \bibinfo {pages} {2567} (\bibinfo {year} {2023})}\BibitemShut {NoStop}%
\bibitem [{\citenamefont {Weissflog}\ \emph {et~al.}(2024)\citenamefont {Weissflog}, \citenamefont {Ma}, \citenamefont {Zhang}, \citenamefont {Fan}, \citenamefont {Lung}, \citenamefont {Pertsch}, \citenamefont {Neshev}, \citenamefont {Saravi}, \citenamefont {Setzpfandt},\ and\ \citenamefont {Sukhorukov}}]{Weissflog_Nanophotonics_2024}%
  \BibitemOpen
  \bibfield  {author} {\bibinfo {author} {\bibfnamefont {M.~A.}\ \bibnamefont {Weissflog}}, \bibinfo {author} {\bibfnamefont {J.}~\bibnamefont {Ma}}, \bibinfo {author} {\bibfnamefont {J.}~\bibnamefont {Zhang}}, \bibinfo {author} {\bibfnamefont {T.}~\bibnamefont {Fan}}, \bibinfo {author} {\bibfnamefont {S.}~\bibnamefont {Lung}}, \bibinfo {author} {\bibfnamefont {T.}~\bibnamefont {Pertsch}}, \bibinfo {author} {\bibfnamefont {D.~N.}\ \bibnamefont {Neshev}}, \bibinfo {author} {\bibfnamefont {S.}~\bibnamefont {Saravi}}, \bibinfo {author} {\bibfnamefont {F.}~\bibnamefont {Setzpfandt}}, \ and\ \bibinfo {author} {\bibfnamefont {A.~A.}\ \bibnamefont {Sukhorukov}},\ }\href {\doibase 10.1515/nanoph-2024-0122} {\bibfield  {journal} {\bibinfo  {journal} {Nanophotonics}\ }\textbf {\bibinfo {volume} {13}},\ \bibinfo {pages} {3563} (\bibinfo {year} {2024})}\BibitemShut {NoStop}%
\bibitem [{\citenamefont {Jia}\ \emph {et~al.}(2024)\citenamefont {Jia}, \citenamefont {Saerens}, \citenamefont {Talts}, \citenamefont {Weigand}, \citenamefont {Chapman}, \citenamefont {Li}, \citenamefont {Grange},\ and\ \citenamefont {Yang}}]{Jia_arXiv_2024}%
  \BibitemOpen
  \bibfield  {author} {\bibinfo {author} {\bibfnamefont {W.}~\bibnamefont {Jia}}, \bibinfo {author} {\bibfnamefont {G.}~\bibnamefont {Saerens}}, \bibinfo {author} {\bibfnamefont {{\"U}.-L.}\ \bibnamefont {Talts}}, \bibinfo {author} {\bibfnamefont {H.}~\bibnamefont {Weigand}}, \bibinfo {author} {\bibfnamefont {R.~J.}\ \bibnamefont {Chapman}}, \bibinfo {author} {\bibfnamefont {L.}~\bibnamefont {Li}}, \bibinfo {author} {\bibfnamefont {R.}~\bibnamefont {Grange}}, \ and\ \bibinfo {author} {\bibfnamefont {Y.}~\bibnamefont {Yang}},\ }\href@noop {} {\bibfield  {journal} {\bibinfo  {journal} {arXiv:2405.03493}\ } (\bibinfo {year} {2024})}\BibitemShut {NoStop}%
\bibitem [{\citenamefont {Ma}\ \emph {et~al.}(2024{\natexlab{a}})\citenamefont {Ma}, \citenamefont {Fan}, \citenamefont {Haggren}, \citenamefont {Molina}, \citenamefont {Parry}, \citenamefont {Shinde}, \citenamefont {Zhang}, \citenamefont {Camacho-Morales}, \citenamefont {Setzpfandt}, \citenamefont {Tan}, \citenamefont {Jagadish}, \citenamefont {Neshev},\ and\ \citenamefont {Sukhorukov}}]{Ma_arxiv_2024b}%
  \BibitemOpen
  \bibfield  {author} {\bibinfo {author} {\bibfnamefont {J.}~\bibnamefont {Ma}}, \bibinfo {author} {\bibfnamefont {T.}~\bibnamefont {Fan}}, \bibinfo {author} {\bibfnamefont {T.}~\bibnamefont {Haggren}}, \bibinfo {author} {\bibfnamefont {L.~V.}\ \bibnamefont {Molina}}, \bibinfo {author} {\bibfnamefont {M.}~\bibnamefont {Parry}}, \bibinfo {author} {\bibfnamefont {S.}~\bibnamefont {Shinde}}, \bibinfo {author} {\bibfnamefont {J.}~\bibnamefont {Zhang}}, \bibinfo {author} {\bibfnamefont {R.}~\bibnamefont {Camacho-Morales}}, \bibinfo {author} {\bibfnamefont {F.}~\bibnamefont {Setzpfandt}}, \bibinfo {author} {\bibfnamefont {H.~H.}\ \bibnamefont {Tan}}, \bibinfo {author} {\bibfnamefont {C.}~\bibnamefont {Jagadish}}, \bibinfo {author} {\bibfnamefont {D.~N.}\ \bibnamefont {Neshev}}, \ and\ \bibinfo {author} {\bibfnamefont {A.~A.}\ \bibnamefont {Sukhorukov}},\ }\href@noop {} {\bibfield  {journal} {\bibinfo  {journal} {arXiv:2409.10845}\ } (\bibinfo {year} {2024}{\natexlab{a}})}\BibitemShut {NoStop}%
\bibitem [{\citenamefont {Li}\ \emph {et~al.}(2020)\citenamefont {Li}, \citenamefont {Liu}, \citenamefont {Ren}, \citenamefont {Wang}, \citenamefont {Su}, \citenamefont {Chen}, \citenamefont {Chu}, \citenamefont {Kuo}, \citenamefont {Liu}, \citenamefont {Zang}, \citenamefont {Guo}, \citenamefont {Zhang}, \citenamefont {Wang}, \citenamefont {Zhu},\ and\ \citenamefont {Tsai}}]{Li_Science_2020}%
  \BibitemOpen
  \bibfield  {author} {\bibinfo {author} {\bibfnamefont {L.}~\bibnamefont {Li}}, \bibinfo {author} {\bibfnamefont {Z.}~\bibnamefont {Liu}}, \bibinfo {author} {\bibfnamefont {X.}~\bibnamefont {Ren}}, \bibinfo {author} {\bibfnamefont {S.}~\bibnamefont {Wang}}, \bibinfo {author} {\bibfnamefont {V.}~\bibnamefont {Su}}, \bibinfo {author} {\bibfnamefont {M.~K.}\ \bibnamefont {Chen}}, \bibinfo {author} {\bibfnamefont {C.~H.}\ \bibnamefont {Chu}}, \bibinfo {author} {\bibfnamefont {H.~Y.}\ \bibnamefont {Kuo}}, \bibinfo {author} {\bibfnamefont {B.}~\bibnamefont {Liu}}, \bibinfo {author} {\bibfnamefont {W.}~\bibnamefont {Zang}}, \bibinfo {author} {\bibfnamefont {G.}~\bibnamefont {Guo}}, \bibinfo {author} {\bibfnamefont {L.}~\bibnamefont {Zhang}}, \bibinfo {author} {\bibfnamefont {Z.}~\bibnamefont {Wang}}, \bibinfo {author} {\bibfnamefont {S.}~\bibnamefont {Zhu}}, \ and\ \bibinfo {author} {\bibfnamefont {D.~P.}\ \bibnamefont {Tsai}},\ }\href {\doibase 10.1126/science.aba9779} {\bibfield  {journal} {\bibinfo  {journal}
  {Science}\ }\textbf {\bibinfo {volume} {368}},\ \bibinfo {pages} {1487–1490} (\bibinfo {year} {2020})}\BibitemShut {NoStop}%
\bibitem [{\citenamefont {Noh}\ \emph {et~al.}(2024)\citenamefont {Noh}, \citenamefont {Santiago-Cruz}, \citenamefont {Sultanov}, \citenamefont {Doiron}, \citenamefont {Gennaro}, \citenamefont {Chekhova},\ and\ \citenamefont {Brener}}]{Noh_NanoLett_2024}%
  \BibitemOpen
  \bibfield  {author} {\bibinfo {author} {\bibfnamefont {J.}~\bibnamefont {Noh}}, \bibinfo {author} {\bibfnamefont {T.}~\bibnamefont {Santiago-Cruz}}, \bibinfo {author} {\bibfnamefont {V.}~\bibnamefont {Sultanov}}, \bibinfo {author} {\bibfnamefont {C.~F.}\ \bibnamefont {Doiron}}, \bibinfo {author} {\bibfnamefont {S.~D.}\ \bibnamefont {Gennaro}}, \bibinfo {author} {\bibfnamefont {M.~V.}\ \bibnamefont {Chekhova}}, \ and\ \bibinfo {author} {\bibfnamefont {I.}~\bibnamefont {Brener}},\ }\href {\doibase 10.1021/acs.nanolett.4c04398} {\bibfield  {journal} {\bibinfo  {journal} {Nano Letters}\ }\textbf {\bibinfo {volume} {24}},\ \bibinfo {pages} {15356–15362} (\bibinfo {year} {2024})}\BibitemShut {NoStop}%
\bibitem [{\citenamefont {Shoji}\ \emph {et~al.}(1997)\citenamefont {Shoji}, \citenamefont {Kondo}, \citenamefont {Kitamoto}, \citenamefont {Shirane},\ and\ \citenamefont {Ito}}]{Shoji_JOSAB_1997}%
  \BibitemOpen
  \bibfield  {author} {\bibinfo {author} {\bibfnamefont {I.}~\bibnamefont {Shoji}}, \bibinfo {author} {\bibfnamefont {T.}~\bibnamefont {Kondo}}, \bibinfo {author} {\bibfnamefont {A.}~\bibnamefont {Kitamoto}}, \bibinfo {author} {\bibfnamefont {M.}~\bibnamefont {Shirane}}, \ and\ \bibinfo {author} {\bibfnamefont {R.}~\bibnamefont {Ito}},\ }\href {\doibase 10.1364/JOSAB.14.002268} {\bibfield  {journal} {\bibinfo  {journal} {J. Opt. Soc. Am. B}\ }\textbf {\bibinfo {volume} {14}},\ \bibinfo {pages} {2268} (\bibinfo {year} {1997})}\BibitemShut {NoStop}%
\bibitem [{\citenamefont {Liu}\ \emph {et~al.}(2018)\citenamefont {Liu}, \citenamefont {Vabishchevich}, \citenamefont {Vaskin}, \citenamefont {Reno}, \citenamefont {Keeler}, \citenamefont {Sinclair}, \citenamefont {Staude},\ and\ \citenamefont {Brener}}]{Liu_NatCommun_2018}%
  \BibitemOpen
  \bibfield  {author} {\bibinfo {author} {\bibfnamefont {S.}~\bibnamefont {Liu}}, \bibinfo {author} {\bibfnamefont {P.~P.}\ \bibnamefont {Vabishchevich}}, \bibinfo {author} {\bibfnamefont {A.}~\bibnamefont {Vaskin}}, \bibinfo {author} {\bibfnamefont {J.~L.}\ \bibnamefont {Reno}}, \bibinfo {author} {\bibfnamefont {G.~A.}\ \bibnamefont {Keeler}}, \bibinfo {author} {\bibfnamefont {M.~B.}\ \bibnamefont {Sinclair}}, \bibinfo {author} {\bibfnamefont {I.}~\bibnamefont {Staude}}, \ and\ \bibinfo {author} {\bibfnamefont {I.}~\bibnamefont {Brener}},\ }\href {\doibase 10.1038/s41467-018-04944-9} {\bibfield  {journal} {\bibinfo  {journal} {Nature Communications}\ }\textbf {\bibinfo {volume} {9}},\ \bibinfo {pages} {2507} (\bibinfo {year} {2018})}\BibitemShut {NoStop}%
\bibitem [{\citenamefont {Sautter}\ \emph {et~al.}(2019)\citenamefont {Sautter}, \citenamefont {Xu}, \citenamefont {Miroshnichenko}, \citenamefont {Lysevych}, \citenamefont {Volkovskaya}, \citenamefont {Smirnova}, \citenamefont {Camacho‐Morales}, \citenamefont {Kamali}, \citenamefont {Karouta}, \citenamefont {Vora}, \citenamefont {Tan}, \citenamefont {Kauranen}, \citenamefont {Staude}, \citenamefont {Jagadish}, \citenamefont {Neshev},\ and\ \citenamefont {Rahmani}}]{Sautter_NanoLett_2019}%
  \BibitemOpen
  \bibfield  {author} {\bibinfo {author} {\bibfnamefont {J.}~\bibnamefont {Sautter}}, \bibinfo {author} {\bibfnamefont {L.}~\bibnamefont {Xu}}, \bibinfo {author} {\bibfnamefont {A.~E.}\ \bibnamefont {Miroshnichenko}}, \bibinfo {author} {\bibfnamefont {M.}~\bibnamefont {Lysevych}}, \bibinfo {author} {\bibfnamefont {I.}~\bibnamefont {Volkovskaya}}, \bibinfo {author} {\bibfnamefont {D.}~\bibnamefont {Smirnova}}, \bibinfo {author} {\bibfnamefont {R.}~\bibnamefont {Camacho‐Morales}}, \bibinfo {author} {\bibfnamefont {K.~Z.}\ \bibnamefont {Kamali}}, \bibinfo {author} {\bibfnamefont {F.}~\bibnamefont {Karouta}}, \bibinfo {author} {\bibfnamefont {K.}~\bibnamefont {Vora}}, \bibinfo {author} {\bibfnamefont {H.~H.}\ \bibnamefont {Tan}}, \bibinfo {author} {\bibfnamefont {M.}~\bibnamefont {Kauranen}}, \bibinfo {author} {\bibfnamefont {I.}~\bibnamefont {Staude}}, \bibinfo {author} {\bibfnamefont {C.}~\bibnamefont {Jagadish}}, \bibinfo {author} {\bibfnamefont {D.~N.}\ \bibnamefont {Neshev}}, \ and\ \bibinfo {author}
  {\bibfnamefont {M.}~\bibnamefont {Rahmani}},\ }\href {\doibase 10.1021/acs.nanolett.9b01112} {\bibfield  {journal} {\bibinfo  {journal} {Nano Letters}\ }\textbf {\bibinfo {volume} {19}},\ \bibinfo {pages} {3905–3911} (\bibinfo {year} {2019})}\BibitemShut {NoStop}%
\bibitem [{\citenamefont {Xu}\ \emph {et~al.}(2020)\citenamefont {Xu}, \citenamefont {Saerens}, \citenamefont {Timofeeva}, \citenamefont {Smirnova}, \citenamefont {Volkovskaya}, \citenamefont {Lysevych}, \citenamefont {Camacho‐Morales}, \citenamefont {Cai}, \citenamefont {Kamali}, \citenamefont {Huang}, \citenamefont {Karouta}, \citenamefont {Tan}, \citenamefont {Jagadish}, \citenamefont {Miroshnichenko}, \citenamefont {Grange}, \citenamefont {Neshev},\ and\ \citenamefont {Rahmani}}]{Xu_ACSNano_2020}%
  \BibitemOpen
  \bibfield  {author} {\bibinfo {author} {\bibfnamefont {L.}~\bibnamefont {Xu}}, \bibinfo {author} {\bibfnamefont {G.}~\bibnamefont {Saerens}}, \bibinfo {author} {\bibfnamefont {M.}~\bibnamefont {Timofeeva}}, \bibinfo {author} {\bibfnamefont {D.}~\bibnamefont {Smirnova}}, \bibinfo {author} {\bibfnamefont {I.}~\bibnamefont {Volkovskaya}}, \bibinfo {author} {\bibfnamefont {M.}~\bibnamefont {Lysevych}}, \bibinfo {author} {\bibfnamefont {R.}~\bibnamefont {Camacho‐Morales}}, \bibinfo {author} {\bibfnamefont {M.}~\bibnamefont {Cai}}, \bibinfo {author} {\bibfnamefont {K.~Z.}\ \bibnamefont {Kamali}}, \bibinfo {author} {\bibfnamefont {L.}~\bibnamefont {Huang}}, \bibinfo {author} {\bibfnamefont {F.}~\bibnamefont {Karouta}}, \bibinfo {author} {\bibfnamefont {H.~H.}\ \bibnamefont {Tan}}, \bibinfo {author} {\bibfnamefont {C.}~\bibnamefont {Jagadish}}, \bibinfo {author} {\bibfnamefont {A.~E.}\ \bibnamefont {Miroshnichenko}}, \bibinfo {author} {\bibfnamefont {R.}~\bibnamefont {Grange}}, \bibinfo {author} {\bibfnamefont
  {D.~N.}\ \bibnamefont {Neshev}}, \ and\ \bibinfo {author} {\bibfnamefont {M.}~\bibnamefont {Rahmani}},\ }\href {\doibase 10.1021/acsnano.9b07117} {\bibfield  {journal} {\bibinfo  {journal} {ACS Nano}\ }\textbf {\bibinfo {volume} {14}},\ \bibinfo {pages} {1379} (\bibinfo {year} {2020})}\BibitemShut {NoStop}%
\bibitem [{\citenamefont {Camacho-Morales}\ \emph {et~al.}(2021)\citenamefont {Camacho-Morales}, \citenamefont {Rocco}, \citenamefont {Xu}, \citenamefont {Gili}, \citenamefont {Dimitrov}, \citenamefont {Stoyanov}, \citenamefont {Ma}, \citenamefont {Komar}, \citenamefont {Lysevych}, \citenamefont {Karouta}, \citenamefont {Dreischuh}, \citenamefont {Tan}, \citenamefont {Leo}, \citenamefont {Angelis}, \citenamefont {Jagadish}, \citenamefont {Miroshnichenko}, \citenamefont {Rahmani},\ and\ \citenamefont {Neshev}}]{Camacho-Morales_AdvPhotonics_2021}%
  \BibitemOpen
  \bibfield  {author} {\bibinfo {author} {\bibfnamefont {R.}~\bibnamefont {Camacho-Morales}}, \bibinfo {author} {\bibfnamefont {D.}~\bibnamefont {Rocco}}, \bibinfo {author} {\bibfnamefont {L.}~\bibnamefont {Xu}}, \bibinfo {author} {\bibfnamefont {V.~F.}\ \bibnamefont {Gili}}, \bibinfo {author} {\bibfnamefont {N.}~\bibnamefont {Dimitrov}}, \bibinfo {author} {\bibfnamefont {L.}~\bibnamefont {Stoyanov}}, \bibinfo {author} {\bibfnamefont {Z.}~\bibnamefont {Ma}}, \bibinfo {author} {\bibfnamefont {A.}~\bibnamefont {Komar}}, \bibinfo {author} {\bibfnamefont {M.}~\bibnamefont {Lysevych}}, \bibinfo {author} {\bibfnamefont {F.}~\bibnamefont {Karouta}}, \bibinfo {author} {\bibfnamefont {A.~A.}\ \bibnamefont {Dreischuh}}, \bibinfo {author} {\bibfnamefont {H.~H.}\ \bibnamefont {Tan}}, \bibinfo {author} {\bibfnamefont {G.}~\bibnamefont {Leo}}, \bibinfo {author} {\bibfnamefont {C.~D.}\ \bibnamefont {Angelis}}, \bibinfo {author} {\bibfnamefont {C.}~\bibnamefont {Jagadish}}, \bibinfo {author} {\bibfnamefont {A.~E.}\
  \bibnamefont {Miroshnichenko}}, \bibinfo {author} {\bibfnamefont {M.}~\bibnamefont {Rahmani}}, \ and\ \bibinfo {author} {\bibfnamefont {D.~N.}\ \bibnamefont {Neshev}},\ }\href {\doibase 10.1117/1.AP.3.3.036002} {\bibfield  {journal} {\bibinfo  {journal} {Advanced Photonics}\ }\textbf {\bibinfo {volume} {3}},\ \bibinfo {pages} {036002} (\bibinfo {year} {2021})}\BibitemShut {NoStop}%
\bibitem [{\citenamefont {Yang}\ \emph {et~al.}(2024)\citenamefont {Yang}, \citenamefont {Weissflog}, \citenamefont {Fedorova}, \citenamefont {Barreda}, \citenamefont {B{\"o}rner}, \citenamefont {Eilenberger}, \citenamefont {Pertsch},\ and\ \citenamefont {Staude}}]{Yang_Nanophotonics_2024}%
  \BibitemOpen
  \bibfield  {author} {\bibinfo {author} {\bibfnamefont {M.}~\bibnamefont {Yang}}, \bibinfo {author} {\bibfnamefont {M.~A.}\ \bibnamefont {Weissflog}}, \bibinfo {author} {\bibfnamefont {Z.}~\bibnamefont {Fedorova}}, \bibinfo {author} {\bibfnamefont {A.~I.}\ \bibnamefont {Barreda}}, \bibinfo {author} {\bibfnamefont {S.}~\bibnamefont {B{\"o}rner}}, \bibinfo {author} {\bibfnamefont {F.}~\bibnamefont {Eilenberger}}, \bibinfo {author} {\bibfnamefont {T.}~\bibnamefont {Pertsch}}, \ and\ \bibinfo {author} {\bibfnamefont {I.}~\bibnamefont {Staude}},\ }\href@noop {} {\bibfield  {journal} {\bibinfo  {journal} {Nanophotonics}\ }\textbf {\bibinfo {volume} {13}},\ \bibinfo {pages} {3311} (\bibinfo {year} {2024})}\BibitemShut {NoStop}%
\bibitem [{\citenamefont {Hsu}\ \emph {et~al.}(2016)\citenamefont {Hsu}, \citenamefont {Zhen}, \citenamefont {Stone}, \citenamefont {Joannopoulos},\ and\ \citenamefont {Solja{\v c}i{\'c}}}]{Hsu_NatMat_2016}%
  \BibitemOpen
  \bibfield  {author} {\bibinfo {author} {\bibfnamefont {C.~W.}\ \bibnamefont {Hsu}}, \bibinfo {author} {\bibfnamefont {B.}~\bibnamefont {Zhen}}, \bibinfo {author} {\bibfnamefont {A.~D.}\ \bibnamefont {Stone}}, \bibinfo {author} {\bibfnamefont {J.~D.}\ \bibnamefont {Joannopoulos}}, \ and\ \bibinfo {author} {\bibfnamefont {M.}~\bibnamefont {Solja{\v c}i{\'c}}},\ }\href@noop {} {\bibfield  {journal} {\bibinfo  {journal} {Nature Reviews Materials}\ }\textbf {\bibinfo {volume} {1}},\ \bibinfo {pages} {16048} (\bibinfo {year} {2016})}\BibitemShut {NoStop}%
\bibitem [{\citenamefont {Koshelev}\ \emph {et~al.}(2018)\citenamefont {Koshelev}, \citenamefont {Lepeshov}, \citenamefont {Liu}, \citenamefont {Bogdanov},\ and\ \citenamefont {Kivshar}}]{Koshelev_PRL_2018}%
  \BibitemOpen
  \bibfield  {author} {\bibinfo {author} {\bibfnamefont {K.}~\bibnamefont {Koshelev}}, \bibinfo {author} {\bibfnamefont {S.}~\bibnamefont {Lepeshov}}, \bibinfo {author} {\bibfnamefont {M.}~\bibnamefont {Liu}}, \bibinfo {author} {\bibfnamefont {A.}~\bibnamefont {Bogdanov}}, \ and\ \bibinfo {author} {\bibfnamefont {Y.}~\bibnamefont {Kivshar}},\ }\href {\doibase 10.1103/PhysRevLett.121.193903} {\bibfield  {journal} {\bibinfo  {journal} {Phys. Rev. Lett.}\ }\textbf {\bibinfo {volume} {121}},\ \bibinfo {pages} {193903} (\bibinfo {year} {2018})}\BibitemShut {NoStop}%
\bibitem [{\citenamefont {Campione}\ \emph {et~al.}(2016)\citenamefont {Campione}, \citenamefont {Liu}, \citenamefont {Basilio}, \citenamefont {Warne}, \citenamefont {Langston}, \citenamefont {Luk}, \citenamefont {Wendt}, \citenamefont {Reno}, \citenamefont {Keeler}, \citenamefont {Brener},\ and\ \citenamefont {Sinclair}}]{Campione_ACSPhoton_2016}%
  \BibitemOpen
  \bibfield  {author} {\bibinfo {author} {\bibfnamefont {S.}~\bibnamefont {Campione}}, \bibinfo {author} {\bibfnamefont {S.}~\bibnamefont {Liu}}, \bibinfo {author} {\bibfnamefont {L.~I.}\ \bibnamefont {Basilio}}, \bibinfo {author} {\bibfnamefont {L.~K.}\ \bibnamefont {Warne}}, \bibinfo {author} {\bibfnamefont {W.~L.}\ \bibnamefont {Langston}}, \bibinfo {author} {\bibfnamefont {T.~S.}\ \bibnamefont {Luk}}, \bibinfo {author} {\bibfnamefont {J.~R.}\ \bibnamefont {Wendt}}, \bibinfo {author} {\bibfnamefont {J.~L.}\ \bibnamefont {Reno}}, \bibinfo {author} {\bibfnamefont {G.~A.}\ \bibnamefont {Keeler}}, \bibinfo {author} {\bibfnamefont {I.}~\bibnamefont {Brener}}, \ and\ \bibinfo {author} {\bibfnamefont {M.~B.}\ \bibnamefont {Sinclair}},\ }\href@noop {} {\bibfield  {journal} {\bibinfo  {journal} {ACS Photonics}\ }\textbf {\bibinfo {volume} {3}},\ \bibinfo {pages} {2362} (\bibinfo {year} {2016})}\BibitemShut {NoStop}%
\bibitem [{\citenamefont {Vabishchevich}\ \emph {et~al.}(2018)\citenamefont {Vabishchevich}, \citenamefont {Liu}, \citenamefont {Sinclair}, \citenamefont {Keeler}, \citenamefont {Peake},\ and\ \citenamefont {Brener}}]{Vabishchevich_ACSPhotonics_2018}%
  \BibitemOpen
  \bibfield  {author} {\bibinfo {author} {\bibfnamefont {P.~P.}\ \bibnamefont {Vabishchevich}}, \bibinfo {author} {\bibfnamefont {S.}~\bibnamefont {Liu}}, \bibinfo {author} {\bibfnamefont {M.~B.}\ \bibnamefont {Sinclair}}, \bibinfo {author} {\bibfnamefont {G.~A.}\ \bibnamefont {Keeler}}, \bibinfo {author} {\bibfnamefont {G.~M.}\ \bibnamefont {Peake}}, \ and\ \bibinfo {author} {\bibfnamefont {I.}~\bibnamefont {Brener}},\ }\href {\doibase 10.1021/acsphotonics.7b01478} {\bibfield  {journal} {\bibinfo  {journal} {ACS Photonics}\ }\textbf {\bibinfo {volume} {5}},\ \bibinfo {pages} {1685–1690} (\bibinfo {year} {2018})}\BibitemShut {NoStop}%
\bibitem [{\citenamefont {Valencia}\ \emph {et~al.}(2002)\citenamefont {Valencia}, \citenamefont {Chekhova}, \citenamefont {Trifonov},\ and\ \citenamefont {Shih}}]{Valencia_PRL_2002}%
  \BibitemOpen
  \bibfield  {author} {\bibinfo {author} {\bibfnamefont {A.}~\bibnamefont {Valencia}}, \bibinfo {author} {\bibfnamefont {M.~V.}\ \bibnamefont {Chekhova}}, \bibinfo {author} {\bibfnamefont {A.}~\bibnamefont {Trifonov}}, \ and\ \bibinfo {author} {\bibfnamefont {Y.}~\bibnamefont {Shih}},\ }\href {\doibase 10.1103/PhysRevLett.88.183601} {\bibfield  {journal} {\bibinfo  {journal} {Phys. Rev. Lett.}\ }\textbf {\bibinfo {volume} {88}},\ \bibinfo {pages} {183601} (\bibinfo {year} {2002})}\BibitemShut {NoStop}%
\bibitem [{\citenamefont {Ma}\ \emph {et~al.}(2023)\citenamefont {Ma}, \citenamefont {Zhang}, \citenamefont {Jiang}, \citenamefont {Fan}, \citenamefont {Parry}, \citenamefont {Neshev},\ and\ \citenamefont {Sukhorukov}}]{MA_NanoLett_2023}%
  \BibitemOpen
  \bibfield  {author} {\bibinfo {author} {\bibfnamefont {J.}~\bibnamefont {Ma}}, \bibinfo {author} {\bibfnamefont {J.}~\bibnamefont {Zhang}}, \bibinfo {author} {\bibfnamefont {Y.}~\bibnamefont {Jiang}}, \bibinfo {author} {\bibfnamefont {T.}~\bibnamefont {Fan}}, \bibinfo {author} {\bibfnamefont {M.}~\bibnamefont {Parry}}, \bibinfo {author} {\bibfnamefont {D.~N.}\ \bibnamefont {Neshev}}, \ and\ \bibinfo {author} {\bibfnamefont {A.~A.}\ \bibnamefont {Sukhorukov}},\ }\href@noop {} {\bibfield  {journal} {\bibinfo  {journal} {Nano Letters}\ }\textbf {\bibinfo {volume} {23}},\ \bibinfo {pages} {8091} (\bibinfo {year} {2023})}\BibitemShut {NoStop}%
\bibitem [{\citenamefont {Ma}\ \emph {et~al.}(2024{\natexlab{b}})\citenamefont {Ma}, \citenamefont {Ren}, \citenamefont {Zhang}, \citenamefont {Meng}, \citenamefont {McManus-Barrett}, \citenamefont {Crozier},\ and\ \citenamefont {Sukhorukov}}]{Ma_arxiv_2024}%
  \BibitemOpen
  \bibfield  {author} {\bibinfo {author} {\bibfnamefont {J.}~\bibnamefont {Ma}}, \bibinfo {author} {\bibfnamefont {J.}~\bibnamefont {Ren}}, \bibinfo {author} {\bibfnamefont {J.}~\bibnamefont {Zhang}}, \bibinfo {author} {\bibfnamefont {J.}~\bibnamefont {Meng}}, \bibinfo {author} {\bibfnamefont {C.}~\bibnamefont {McManus-Barrett}}, \bibinfo {author} {\bibfnamefont {K.~B.}\ \bibnamefont {Crozier}}, \ and\ \bibinfo {author} {\bibfnamefont {A.~A.}\ \bibnamefont {Sukhorukov}},\ }\href@noop {} {\bibfield  {journal} {\bibinfo  {journal} {arXiv:2408.02903}\ } (\bibinfo {year} {2024}{\natexlab{b}})}\BibitemShut {NoStop}%
\bibitem [{\citenamefont {Santiago-Cruz}\ \emph {et~al.}(2021{\natexlab{b}})\citenamefont {Santiago-Cruz}, \citenamefont {Sultanov}, \citenamefont {Zhang}, \citenamefont {Krivitsky},\ and\ \citenamefont {Chekhova}}]{Santiago-Cruz_OptLett_2021}%
  \BibitemOpen
  \bibfield  {author} {\bibinfo {author} {\bibfnamefont {T.}~\bibnamefont {Santiago-Cruz}}, \bibinfo {author} {\bibfnamefont {V.}~\bibnamefont {Sultanov}}, \bibinfo {author} {\bibfnamefont {H.}~\bibnamefont {Zhang}}, \bibinfo {author} {\bibfnamefont {L.~A.}\ \bibnamefont {Krivitsky}}, \ and\ \bibinfo {author} {\bibfnamefont {M.~V.}\ \bibnamefont {Chekhova}},\ }\href {\doibase 10.1364/OL.411176} {\bibfield  {journal} {\bibinfo  {journal} {Optics Letters}\ }\textbf {\bibinfo {volume} {46}},\ \bibinfo {pages} {653} (\bibinfo {year} {2021}{\natexlab{b}})}\BibitemShut {NoStop}%
\bibitem [{\citenamefont {Gennaro}\ \emph {et~al.}(2022)\citenamefont {Gennaro}, \citenamefont {Doiron}, \citenamefont {Karl}, \citenamefont {Iyer}, \citenamefont {Serkland}, \citenamefont {Sinclair},\ and\ \citenamefont {Brener}}]{Gennaro_ACSPhotonics_2022}%
  \BibitemOpen
  \bibfield  {author} {\bibinfo {author} {\bibfnamefont {S.~D.}\ \bibnamefont {Gennaro}}, \bibinfo {author} {\bibfnamefont {C.~F.}\ \bibnamefont {Doiron}}, \bibinfo {author} {\bibfnamefont {N.}~\bibnamefont {Karl}}, \bibinfo {author} {\bibfnamefont {P.~P.}\ \bibnamefont {Iyer}}, \bibinfo {author} {\bibfnamefont {D.~K.}\ \bibnamefont {Serkland}}, \bibinfo {author} {\bibfnamefont {M.~B.}\ \bibnamefont {Sinclair}}, \ and\ \bibinfo {author} {\bibfnamefont {I.}~\bibnamefont {Brener}},\ }\href {\doibase 10.1021/acsphotonics.1c01937} {\bibfield  {journal} {\bibinfo  {journal} {ACS Photonics}\ }\textbf {\bibinfo {volume} {9}},\ \bibinfo {pages} {1026–1032} (\bibinfo {year} {2022})}\BibitemShut {NoStop}%
\end{thebibliography}

%

\end{document}